\documentclass{article} 

\usepackage{amsmath}
\usepackage{url,hyperref,lineno,microtype,subcaption}
\usepackage[margin=3cm]{geometry}
\usepackage{authblk}
\usepackage{amssymb}
\usepackage{multicol}
\usepackage{graphicx}
\usepackage{color, colortbl}
\usepackage{color,soul}
\usepackage{tikz}
\usetikzlibrary{automata, positioning}
\usepackage{tcolorbox}
\usepackage{tabularx}
\usepackage{array}
\usepackage{colortbl}
\tcbuselibrary{skins}

\newcolumntype{Y}{>{\raggedleft\arraybackslash}X}

\tcbset{tab1/.style={fonttitle=\bfseries\large,fontupper=\normalsize\sffamily,
colback=yellow!10!white,colframe=red!75!black,colbacktitle=Salmon!40!white,
coltitle=black,center title,freelance,frame code={
\foreach \n in {north east,north west,south east,south west}
{\path [fill=red!75!black] (interior.\n) circle (3mm); };},}}

\tcbset{tab2/.style={enhanced,fonttitle=\bfseries,fontupper=\normalsize\sffamily,
colback=yellow!10!white,colframe=red!50!black,colbacktitle=white!40!white,
coltitle=black,center title}}

\title{\textbf{Comparisons of wave dynamics in Hodgkin-Huxley and Markov-state
formalisms for the Sodium (Na) channel in some mathematical models for human cardiac
tissue}} 

\author{Mahesh Kumar Mulimani\ $^{1}$}
\author{Alok Ranjan Nayak\,$^{2}$}
\author{Rahul Pandit,$^{1,}$\thanks{Also at Jawaharlal Nehru Centre For Advanced Scientific Research,
Jakkur, Bangalore, India; rahul@iisc.ac.in } }
\affil{$^{1}$Centre for Condensed Matter Theory, Department of Physics, Indian 
Institute of Science, Bangalore 560012, India. \newline
$^{2}$International Institute of Information and Technology, Bhubaneshwar, Orissa, India.}
\date{}
\begin{document}
\onecolumn
\maketitle

\begin{abstract}

We compare and contrast spiral- and scroll-wave dynamics in five different
mathematical models for cardiac tissue.  The first is the TP06 model, due to ten
Tusscher and Panfilov~\cite{ten2006alternans}, which is based on the
Hodgkin-Huxley formalism; the remaining four are Markov-state models, MM1 WT
and MM2 WT, for the wild-type (WT) Na channel, and MM1 MUT and MM2 MUT, for
the mutant Na channel~\cite{vecchietti2006computer,moreno2011computational}.
Our results are based on extensive direct numerical simulations of waves of
electrical activation in these models, in two- and three-dimensional (2D and
3D) homogeneous simulation domains and also in domains with localised heterogeneities,
either obstacles with randomly distributed inexcitable regions or mutant cells
in a wild-type background. Our study brings out the sensitive dependence of
spiral- and scroll-wave dynamics on these five models and the parameters that
define them. We also explore the control of spiral-wave turbulence in these
models.

\small
\section*{Keywords:}  Mathematical Models for Cardiac Tissue,
Hodgkin-Huxley Models, Markov-state Models, 
Wild-Type Markov Models, Mutant Markov Models.

\end{abstract}

\section{Introduction}

The development of an understanding of the dynamics of waves of electrical
activation in cardiac tissue is a problem of central importance in research on
life-threatening cardiac arrhythmias, because sudden cardiac death is responsible
for roughly half of the deaths because of cardiovascular disease, i.e., $15\%$
of all deaths globally ~\cite{mehra2007global}. Approximately  $80\%$ of sudden
cardiac deaths arise from ventricular arrhythmias ~\cite{mehra2007global}. Such
arrhythmias are often associated with the formation of spiral or scroll waves
of electrical activation; unbroken spirals or scrolls lead to ventricular
tachycardia (VT), whereas broken waves, with spiral- or scroll-wave turbulence
~\cite{kleber2004basic,shajahan2007spiral,shajahan2009spiral,clayton2011models,majumder2011scroll,majumder2011overview},
are responsible for ventricular fibrillation (VF); VT and VF lead to
the malfunctioning of the pumping mechanism of the heart, so, in the absence of
medical intervention, VF leads to sudden cardiac death. It is very important,
therefore, to study VT and VF by using all means possible, namely, \textit{in
vivo}, \textit{in vitro}, and \textit{in silico} investigations, which play
complementary roles. \textit{In silico} investigations require mathematical
models for cardiac cells (cardiomyocytes or, simply, myocytes) and for cardiac
tissue.

Studies of the electrical behavior of myocytes require models for the dynamics
of the ion
channels~\cite{kleber2004basic,shajahan2009spiral,clayton2011models,c2b57e77bd614157b764bd42b8c6f47f}.
The first, successful ion-channel model, due to Hodgkin and
Huxley~\cite{hodgkin1952quantitative}, considers the opening and closing of the
channel to be governed by the gates, which depend, in turn, on the myocyte
transmembrane potential $V_m$; these Hodgkin-Huxley-Model (HHM) gates are
defined by  deterministic, first-order, ordinary differential Equations (ODEs);
and each gating variable is independent of other gating variables. However,
ion channels are proteins that can have many conformational states and,
therefore, channels open and close stochastically; hence, discrete-state Markov
models (MMs) have been developed to model ion channels; in some cases, these
Markov models can be reduced to HHMs~\cite{keener2009invariant}. Clearly, these
Markov models are more general than HHMs; in particular, the discrete states in
an MM depend on each other; and MMs have more parameters than HHMs.  

Markov-state models are especially useful when there are ion-channel mutations
in which the functionality of an ion-channel subunit is disturbed. For example,
mutations in the HERG subunit in the  rapid, delayed, rectifier potassium (Kr)
channel lead to the prolongation of the myocyte action-potential duration
(APD); this is referred to as the LQT2 syndrome~\cite{clancy2001cellular}; and
mutations in the $\alpha$-subunit in the Na channel result in the LQT3
syndrome~\cite{clancy1999linking}, which can lead to sudden cardiac death. The
MM formalism has been used to study the effects of mutations in a variety of
ion channels~\cite{fink2009markov} and especially on the LQT syndrome because of
mutations in the Kr channel~\cite{clancy2001cellular} and in the Na
channel~\cite{moreno2011computational,clancy1999linking}. In particular, such
studies have elucidated the effects of different mutations on the myocyte
action potential (AP)~\cite{clancy1999linking} and the interaction between
drugs and the discrete states in an
MM~\cite{moreno2011computational,moreno2013ranolazine,clancy2007pharmacogenetics}.
Challenges in MM studies include the difficulties in estimating the large
number of parameters in these models~\cite{fink2009markov} and the higher
computational cost relative to HHM investigations.

Recently it has been shown that, at the level of a single cardiomyocyte, the
dynamics of wild-type (WT) and mutant (MUT) ion-channels can be modeled well by
the HHM formalism, if it is obtained from the Markov-state Model
(MM)~\cite{carbonell2016comparison}.  In particular, HHM action potentials,
their morphological properties, the action-potential-duration restitution
(APDR), and the conduction-velocity restitution (CVR) are comparable to
their MM counterparts~\cite{carbonell2016comparison}. The authors
of Ref.~\cite{carbonell2016comparison} have considered both WT and MUT cases
for Kr and Na channels in the MM and their HHM counterparts; their results are
encouraging, insofar as they suggest that we can use simple, effective HHM
models, whose parameters are obtained from comparisons with their complicated
MM counterparts, to obtain the properties of action potentials and their
dependence on mutations. A careful comparison of these Markov models and the
Hodgkin-Huxley model for an ion channel, at the cellular level, brings out the
differences in the action potential and its morphology.  To compare the
characteristic properties of excitation waves in these models, it behooves us
to carry out studies of spiral- and scroll-wave dynamics in homogeneous and
heterogeneous tissue in two- and three-dimensional (2D and 3D, respectively)
simulation domains; we embark on such a study here. In particular, we focus on
the Na channels in these MM and HHM models, as the Na channel is important in
controling the upstroke-velocity, at the cellular level, and CV, at the tissue level. 

In our Hodgkin-Huxley model (HHM) Na-channel formalism, we use the
human-ventricular-tissue TP06 model, due to ten Tusscher and
Panfilov~\cite{ten2006alternans}. We compare spiral- and scroll-wave states in
this model with their counterparts in two different Markov-state models, which
we call MM1~\cite{vecchietti2006computer} and
MM2~\cite{moreno2011computational}. In these models, we study both wild-type
(WT) and mutant (MUT) Na channels, by replacing the Na-channel formalism in the
TP06 model by their MM1 and MM2 versions; and we use the TP06 formulation for
all other ion channels (see the section on Methods). Therefore, we examine
three models for the WT Na channel (these are variants of the TP06 model): the
original HHM (TP06) and two MMs (MM1 WT and MM2 WT); and we use two models for
the mutant channels (again variants of the TP06 model), specifically, the MM1
MUT and MM2 MUT models. Note that the TP06 HHM is not obtained from the Markov
models as in Ref.~\cite{carbonell2016comparison}. 

We first compare activation and inactivation properties of the Na channels in
all the five models mentioned above. We then contrast the effects of these
changes on the action potentials and their morphologies in these models, at the
single-cell level. We show that, for the wild-type (WT) Na-channel, the
probability of opening of this channel is different for the TP06, MM1 WT, and
MM2 WT models. The peak value of this probability and the time duration of this
opening are also dissimilar in MM1 WT and MM2 WT models. These differences
alter the action potential (AP) and its morphology.  We show that, for the
mutant (MUT) Na channels, the failure of inactivation leads to early
afterdepolarizations (EADs), in the APs in MM1 MUT and MM2 MUT
models~\cite{zimik2015comparative}.

The differences in the WT Na peak amplitude lead to disparate upstroke
velocities in these models, which manifest themselves in dissimilar CVRs.
Furthermore, the conduction velocities (CVs) in MM1 WT and MM2 WT models turn
out to be outside (lower than) the accepted range for CV in the human
myocardium; we show that we can obtain CVs in this range if we increase the
diffusion constant $D$ in both MM1 WT and MM2 WT models. The differences in our
single-cell and cable-level results motivate our study of wave dynamics in
mathematical models for cardiac tissue, which use these different models.

We carry out a variety of simulations in 2D homogeneous domains to show that
spiral-wave dynamics, in TP06, MM1 WT, MM2 WT, MM1 MUT, and MM2 MUT models,
depends sensitively on these models. For example, we demonstrate that, in the
MM1 WT (MM2 WT) model, the spiral wave is stable (unstable, meandering spiral).
The formation of certain EADs can lead to backward propagation of the wave, and
rapid spiral breakup, in the MM2 MUT model; by contrast, in the MM1 MUT model,
EADs are somewhat different, so we do not find such backward propagation, the
mother rotor is unaffected, and there is only far-field breakup of the spiral.
Furthermore, the spatiotemporal evolution of a spiral wave in the MM2 WT model
depends sensitively on the time $\tau_{S2}$ between the application of the S1
and S2 impulses, which we use to initiate spiral waves.

In the case of mutant models, because of the different kinds of
EADs that we find in MM1 MUT and MM2 MUT, these models display qualitatively
different electrical-wave dynamics. Furthermore, in the spirit of the studies
of Refs.~\cite{shajahan2007spiral,shajahan2009spiral,majumder2014turbulent,zimik2017reentry},
we investigate the effects of two types of inhomogenieties on spiral-wave
dynamics in these models: (a) Two-dimensional (2D), circular or
three-dimensional (3D), cylindrical obstacles, with a random distribution of
inexcitable regions, to model fibrotic patches in Markov-state WT models;
$P_{f}$, the percentage of inexcitable obstacles, and the radius of the
obstacle are important control parameters. (b) A circular patch of mutant cells
in an otherwise homogeneous, 2D WT domain; we find that a spiral wave is formed
in the MM2 MUT model, but not in the MM1 MUT model, if we pace the tissue at a
high frequency.

The elimination of spiral- and scroll-wave turbulence is of central importance
in developing low-amplitude defibrillation schemes for the elimination of VT
and VF. In the Supplementary Material, we describe one such defibrillation scheme (control of spiral waves) for the models of we study.

We carry out a few illustrative studies of scroll waves in 3D TP06, MM1 WT, and
MM2 WT models.  We show, in a homogeneous domain, that scroll waves are stable in TP06 and MM1 WT models, but not in the MM2 model. 
We also investigate when scroll waves are anchored or broken up by
cylindrical obstacles, of the type described above.

Finally, we perform a parameter-sensitivity analysis for
TP06 and MM1 WT and MM2 WT models, in which we consider three important dependent
variables, namely, the APD, $V_{max}$, and $V_{rest}$, at the cellular
level and CV and APD at the cable level (see Supplementary Material).

The remaining part of this paper is organized as follows. Section 2 is devoted
to \textbf{{Methods and Simulations}}. In Section 3 we report our
\textbf{{Results}} for single-cell studies and tissue-level simulations in 2D
square and 3D slab domains for WT models; we also present, for MUT models,
single-cell and 2D-simulation results.  In Section 4 , \textbf{{Discussion and
Conclusions}}, we end with concluding remarks.

\section{Methods and Simulations}
\label{sec:Methods}
\subsection{Model}

The electrical behavior of a single cardiac myocyte is governed by the
following ordinary differential Equation (ODE) for the transmembrane potential
$V_{m}$:
\begin{eqnarray}	 
\frac{dV_{m}}{dt} &=& -\frac{I_{ion}}{C_{m}}; \label{eq:Vm}\\
I_{ion} &=& \sum_{i} I_{i};\label{eq:Iion} 
\end{eqnarray}
here, $I_{ion}$ is the sum of all the ionic currents, $I_{i}$ is the current
because of the $i^{th}$ ion-channel, and $C_m$ is the normalized, transmembrane
capacitance. In the parent TP06 model, $I_{ion}$ is the sum of the following
$12$ ionic currents (Table 1):
\begin{equation}
I_{ion} = I_{Na} + I_{CaL}+I_{to}+I_{Ks}+I_{Kr}+I_{K1}+I_{NaCa}+I_{NaK}+I_{pCa}+I_{pK}+I_{bNa}+I_{bCa}.\label{eq:Iionsum} 
\end{equation}
The spatiotemporal evolution of $V_m$, at the tissue level, is governed by the
following reaction-diffusion partial differential Equation (PDE):
\begin{equation}
\frac{\partial{V_{m}}}{\partial{t}} = D \ {\nabla^2}V - \frac{I_{ion}}{C_{m}} ,
	\label{eq:VmPDE}
\end{equation}
where $D$ is the diffusion constant; for simplicity we consider the 
case in which $D$ is a scalar. 

 \subsection*{Hodgkin-Huxley Model}
\label{subsec:HHM}

The TP06 model uses the Hodgkin-Huxley formalism for the WT Na channel. The
macroscopic current through this channel is governed by the three gating
variables $m, h,$ and $j$~\cite{ten2006alternans}; the first of these is an
activation gate and the latter two are inactivation gates; the gating dynamics
and the Na current are given by  
\begin{eqnarray}
\frac{da_n}{dt} &=& \frac{a_{\infty} - a_n}{\tau_n} ; 
\label{eq:TP06Na}  
	\text{here} \ a_{n} \, \text{can be} \ m, h, \text{or} \ j;  \\ 
		I_{Na} &=& G_{Na} \  m^3 h j (V_m-E_{Na}); 
\label{eq:INa} 
\end{eqnarray}
$G_{Na}$ is the maximal sodium-channel conductance, $a_{\infty}$ is 
the steady-state value of $a_n$, $\tau_{n}$ the time constant of this 
gating variable, and $E_{Na}$ is the sodium-channel Nernst potential.

\subsection*{Markov-state models}
\label{MMs}

We consider four Markov-state models (MMs): two of these are for the wild-type
(WT) and the other two for the mutant (MUT) Na channels. We use the
Markov-state formalisms of Ref.~\cite{vecchietti2006computer} for the first WT
and MUT Na channels; we refer to these as MM1 WT and MM1 MUT, respectively. We
use the Markov models of Ref.~\cite{moreno2011computational} for the second WT
and MUT Na channels, which we label MM2 WT and MM2 MUT, respectively.  We then
replace the Na current in the TP06 model by these two different WT and two
different MUT models. Finally, we have three different WT models, i.e., the
original TP06, MM1 WT, and MM2 WT; and we have two different MUT models,
namely, MM1 MUT and MM2 MUT. All the other currents in the original TP06 model
are unaltered in our studies below.

\begin{figure}
\resizebox{\linewidth}{!}{
	\includegraphics[scale=0.45]{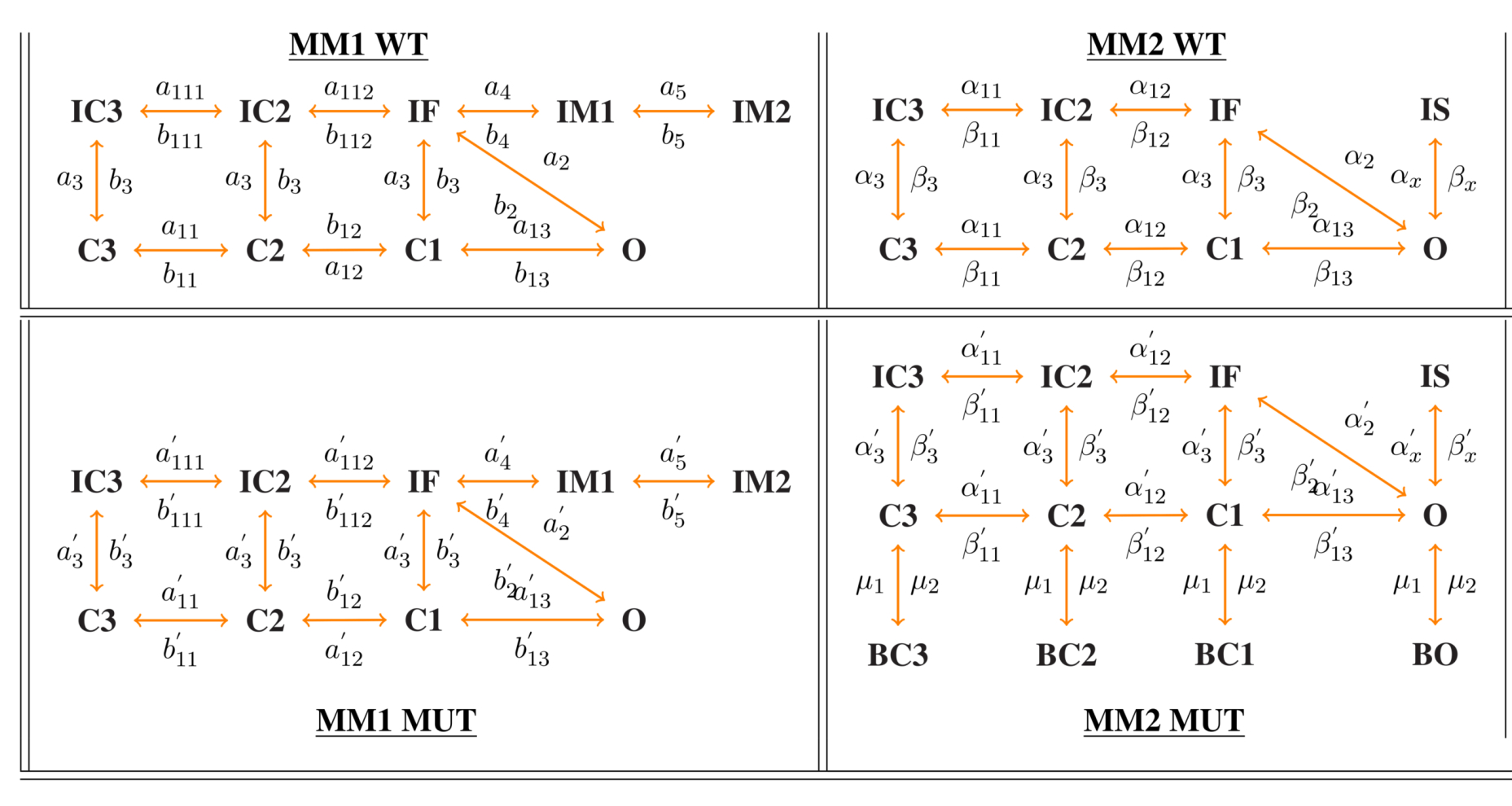}}
 
\caption{\textbf{Schematic diagrams for Markov Models for Wild-Type (WT) and
Mutant (MUT) cases:} Top panel: MM1 WT and MM2 WT models; the MM1 WT model has
$9$ states, namely, the open state ($\bf{O}$), the closed states ($\bf{C1, C2,
C3}$), and the inactivation states ($\bf{IF, IM1, IM2, IC2, IC3}$);  the MM2 WT
model has $8$ states, namely, the open state ($\bf{O}$), the closed states
($\bf{C1, C2, C3}$), and the inactivation states ($\bf{IF, IS, IC2, IC3}$).
The orange, double-headed arrows indicate transitions between such Markov
states; transition rates for the rightward (leftward) transition are given
above (below) these arrows, e.g., $a_{111}$ ($b_{111}$) for the $\bf{IC3} \to
\bf{IC2}$ ($\bf{IC2} \to \bf{IC3}$) transition in MM1 WT. Bottom panel:
schematic diagrams for the MM1 MUT and MM2 MUT modes; the MM1 MUT model has the
same number of states as the MM1 WT model, but the transition rates between the
Markov states are different; the MM2 MUT model has $12$ states: $8$ of these
are as in the MM2 WT model; in addition there are $4$ bursting states, namely,
$\bf {BO, BC1, BC2, BC3}$.}
\label{fig:MM}

\end{figure}

Schematic diagrams of MM1 WT and MM2 WT models are shown in the top panel of
Figure~\ref{fig:MM}. The MM1 WT model has nine states: the open state
($\bf{O}$), the three closed states ($\bf{C1, C2, C3}$), and the five
inactivation states ($\bf{IF, IM1, IM2, IC2, IC3}$).  The MM2 WT  model has
eight states: the open state ($\bf{O}$), the three closed states ($\bf{C1, C2,
C3}$), and the four inactivation states ($\bf{IF, IS, IC2, IC3}$).  The orange,
double-headed arrows indicate transitions between such Markov states;
transition rates for the rightward (leftward) transition are given above
(below) these arrows, e.g., $a_{111}$ ($b_{111}$) for the $\bf{IC3} \to
\bf{IC2}$ ($\bf{IC2} \to \bf{IC3}$) transition in MM1 WT. Similar schematic
diagrams for the MM1 MUT and MM2 MUT models are given in the bottom panel of
Figure~\ref{fig:MM}.  The MM1 MUT model has the same number of states as the
MM1 WT model, but the transition rates between the Markov states are different.
The MM2 MUT model has $12$ states: $8$ of these are as in the MM2 WT model; in
addition, there are $4$ bursting states, namely, $\bf{BO, BC1, BC2, BC3}$.

The dynamics of transitions between the states of these Markov models and the
Na-channel current $I_{Na}$ are given, respectively, by Equations~\ref{eq:dpdt}
and \ref{eq:MMNa} below: 
\begin{equation}
\frac{dP_{k}}{dt} = \sum_{l,l\rightarrow k} \alpha_{l} P_{l} -
\sum_{l,k\rightarrow l} \beta_{l} P_{k}, 
\label{eq:dpdt}
\end{equation}
where $l,k$ label the states $\bf {O,
C1,C2,C3,IC2,IC3,IM1,IM2,IF,IS,BO,BC1,BC2,BC3}$ (Figure~\ref{fig:MM});
$\alpha_{l}$ and $\beta_{l}$ are generic labels for forward and backward
transition rates, respectively;
\begin{equation}
I_{Na} = G_{Na} (P_{O}) (V_{m}-E_{Na}),
\label{eq:MMNa}
\end{equation}
where $P_{O} $ is the probability of the opening of the Na channel. 

We use the values of $G_{Na}$ that are employed either in the original TP06 
model~\cite{ten2006alternans} or in Ref.~\cite{vecchietti2006computer};
specifically, we use
    $$G_{Na} =
    \begin{cases}
    14.838 \ nS/pF , \ \ \text{TP06, MM2 WT, and MM2 MUT;} & \\
    16 \ nS/pF , \ \ \text{MM1 WT and MM1 MUT.} \\
    \end{cases}$$
    \label{Methods:MM}    

The major differences between the two MM (WT and MUT) models are as follows;
\begin{itemize}
\item  The number of states and the connections between them (see Figure~\ref{fig:MM}  
\item The MM2 WT model has a Na current with a late component; this is absent in the MM1 WT model.
\end{itemize}
We show below that these differences can have significant effects on single-cell and spiral-wave properties in these models.

	\subsection{Numerical Simulations}

For our single-cell simulations, we use the Rush-Larsen method to solve
Equation~\ref{eq:TP06Na} for the HHM; for the MM Equation~\ref{eq:dpdt} we use
the implicit trapezoidal method of Ref.~\cite{moreno2011computational} and the
forward-Euler method for Equation~\ref{eq:Vm}.  In our 2D tissue simulations
for Equation~\ref{eq:VmPDE}, we use a square domain with $ N \times N$ grid
points, a fixed space step of size  $\Delta x = 0.025 \ cm $, and a time step
$\Delta t = 0.02 \ ms $; in 3D we use a slab domain (see below). The accuracy of the numerical scheme is tested and is reported in Supplementary Material.For the
homogeneous tissue we consider, $D$ is a scalar; we use $D = 0.00154 \
{cm^2}/{ms} $, as in Ref.~\cite{ten2006alternans}; this yields a maximum
plane-wave conduction velocity $70 {cm}/{s}$, which is in the biophysically
reasonable range for human ventricular tissue ~\cite{ten2004model}. For the
Laplacian in Equation~\ref{eq:VmPDE}, we use five-point and seven-point
stencils, respectively, in 2D and 3D.  We impose no-flux boundary conditions.
When we study the effects of heterogeneities, we introduce, in a localized
region of our simulation domain, a patch of inexcitable cells, which are
decoupled from adjoining cells (effectively, $D = 0$ in this patch).

To obtain the inactivation and activation properties of MM1 WT, MM2 WT, MM1
MUT, and MM2 MUT Na channels, we use the voltage-clamp-simulation protocol of
Ref.~\cite{vecchietti2006computer}. In the activation protocol, we clamp the
cell with a voltage $V_c$, which ranges from the hyper-polarized regime (below
the resting membrane potential of an action potential) to the depolarized
regime ($\simeq 50$ mV); we do this in steps of $5$ mV; the clamping is
maintained for a clamping time $t_c = 1 s$. We then record the peak current
$I_{peak Na}(V_{c})$ and divide it by the driving force $(V-E_{Na})$ to obtain
the conductance $G(V_{c})$, which we normalize to obtain the activation
variable $\mathcal{A}$ as follows:
\begin{eqnarray}
V_{c}  &=& {-100 \ mV \ to \ 50 \ mV} , \ t = 1s ; \nonumber  \\
I_{peak Na}(V_{c}) &=& \text{max} \ ( P_{O} )\ (V_{c}-E_{Na});\nonumber \\ 
G (V_{c}) &=& \frac{I_{peak Na}}{(V-E_{Na})}; \nonumber \\
\mathcal{A} &\equiv& \frac{G (V_{c})}{G(V_{c} = 50 \ mV)}.
\label{eq:act}
\end{eqnarray}

Similarly, for the inactivation protocol, we use a holding potential ($V_h$),
ranging between hyper-polarized and the depolarized values, and apply it for
$\simeq 250 \ ms$; we then apply a test potential $V_{t} = 0 \ mV$, record the
peak Na current, and then define the inactivation variable $\mathcal{I}$ as
follows:
\newline

    $$V_{c} =               
   \begin{cases} 
      V_{h}= {-130 \ mV \ \text{to} \ -10 \ mV} , \ \text{t} < 250 \ ms; &    \\
       V_{t}= 0 \ mV, \ \text{t} = 250 \ ms; &     \\
    \end{cases}$$
$$ I_{peak Na}(V_{c}) = \text{max} \ ( P_{O} ) \ (V_{c}-E_{na}) , \ t > 250 \ ms; $$
    \begin{equation}
    \mathcal{I} \equiv \frac{I_{peak \ Na} (V_{c})}{I_{peak \ Na}(V_{c} = -130 \ mV)}.
    \label{eq:inact}
    \end{equation}

We record the single-cell AP and its morphology, after we have paced the cell
with $n$ pulses, each with a constant pacing cycle length (PCL); we use $n =
500$ pulses. We obtain the static action potential duration restitution
(APDR) (s1-s2) as follows: (a) we apply several pulses ($\simeq 17$) with a
fixed PCL (s1); then, once the system reaches the steady state, we change the
diastolic interval (DI) (s2) by recording the time at which the cell is $90 \%
$ repolarized; this is the action potential duration ($\text{APD}_{90}$), or
simply the APD.  We next obtain the dynamic conduction velocity restitution
(CVR) (s1-s1) as follows: We consider a cable of cells 
(of dimension $832 \times 10$) 
and pace it by applying a current stimulus (s1) at one of its ends; we obtain CV 
from the time that an iso-potential line takes to move between two cells, which
are separated by a fixed distance.  The CVR is the plot of CV versus
DI. 

In our 2D studies we use two representative square domains, namely, one with
$1024 \times 1024 $ grid points, for our spiral-wave studies, and another with
$512 \times 512$ grid points, when we pace of the simulation domain along an
edge. In our studies of scroll-wave dynamics we use a 3D slab domain with $1024
\times 1024 \times 40$ grid points ( $25.6 cm\times 25.6 cm \times 1 cm$).  We
initiate spiral and scroll waves in such domains by using the following S1-S2
cross-field protocol, with stimuli amplitudes of $150 \ pA/pF$ and durations of
$3 \ ms$: We allow a plane wave (S1) to propagate in the domain along a
particular direction; as it propagates, we start another plane wave (S2), in a
direction perpendicular to the S1 wave; this results in a conduction block and,
eventually, the formation of a spiral wave (2D) or scroll wave (3D). We 
vary the time interval  $\tau_{S2}$ between the S1 and S2 impulses to
study the sensitive dependence of the spiral-wave dynamics on  $\tau_{S2}$.

We have carried out the following two sets of simulations of electrical-wave
dynamics with certain localised inhomogeneities in an otherwise homogeneous
simulation domain. (a) In the first set of simulations we introduce circular
(2D) and cylindrical (3D) regions with a random distribution of inexcitable
obstacles to mimic localised fibrotic patches in Markov-state WT models; an
important control parameter here is $P_{f}$, the percentage of inexcitable
obstacles in the circular or cylindrical regions. (b) In the second set of
simulations, we examine electrical-wave dynamics in the presence of a circular
patch of mutant cells in an otherwise homogeneous, 2D WT domain.

\label{Methods:Numscheme}

\section{Results}

\subsection{Single-cell results}

\subsection*{Activation and Inactivation}
\label{sec:actinact}

\begin{figure}
    \resizebox{\linewidth}{!}{
	\includegraphics[scale=0.45]{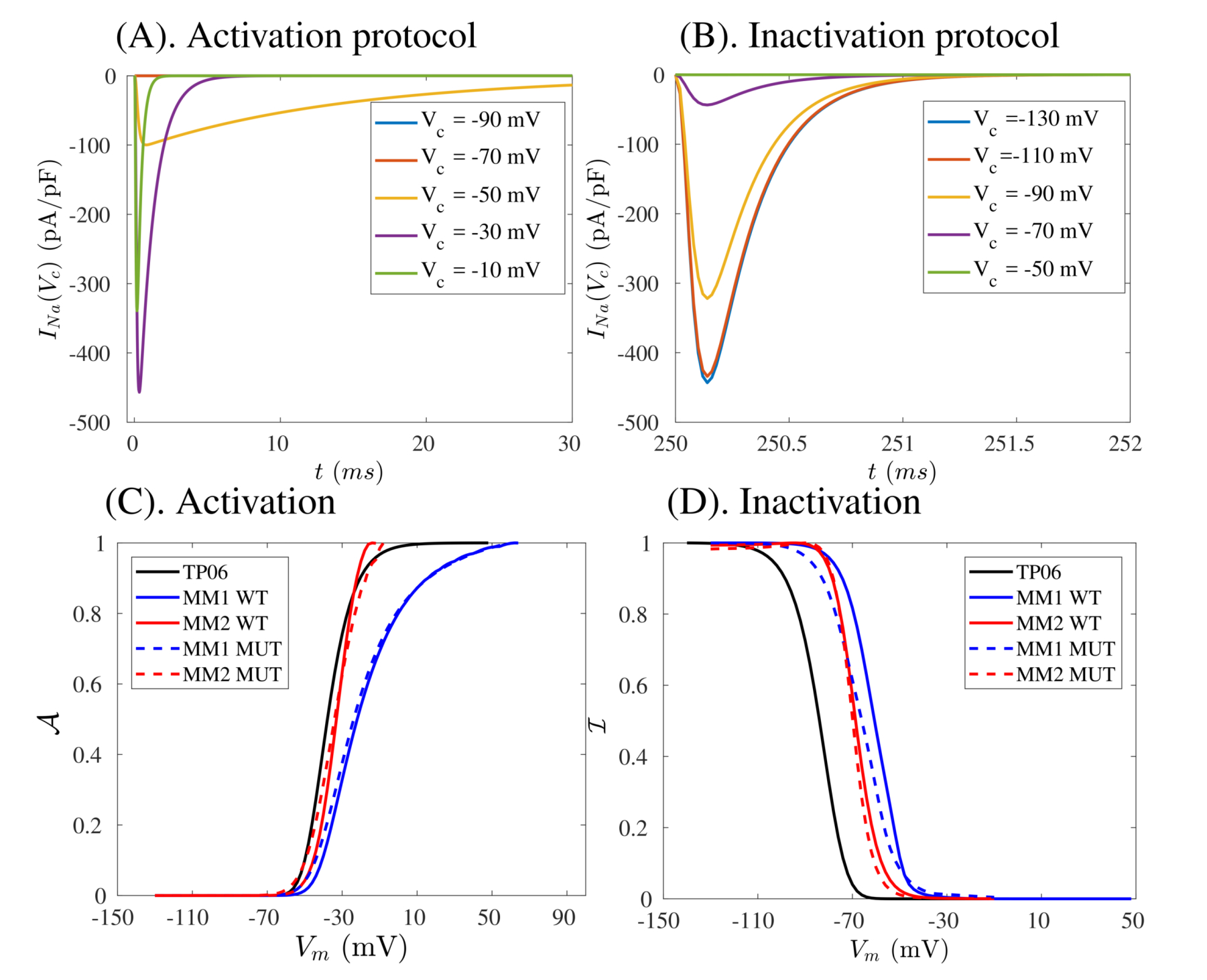}}
	
	\caption{ \textbf{Na-channel activation $\mathcal{A}$ and the
inactivation $\mathcal{I}$, in the models MM1 (WT and MUT) and MM2 (WT and
MUT), compared with their TP06 counterparts.} (A) Activation- and (B)
inactivation-protocol plots, for different values of $V_c$, of $I_{Na}$ versus
time $t$ (see ~\ref{Methods:MM}), whence we obtain the plots of (C) and (D),
which depict, respectively, the dependences of $\mathcal{A}$ and $\mathcal{I}$
on $V_m$, for the MM1 WT, MM2 WT, MM1 MUT, MM2 MUT, and TP06 models; for the
TP06 model $\mathcal{A} \ = m_{\infty}^3$ and $\mathcal{I} \ = \ j_{\infty}
\times h_{\infty}$.}

\label{fig:actinact}
\end{figure}

We begin with a comparison of the activation $\mathcal{A}$ and the inactivation
$\mathcal{I}$ in the models MM1 (WT and MUT) and MM2 (WT and MUT) with their
TP06 counterparts. Figures \ref{fig:actinact}A and \ref{fig:actinact}B show,
for different values of $V_c$, plots of $I_{Na}$ versus time $t$. These plots
illustrate, respectively, the activation and inactivation protocols (see
~\ref{Methods:MM}), whence we obtain Figures \ref{fig:actinact}C and
\ref{fig:actinact}D, which depict, respectively, the dependences of
$\mathcal{A}$ and $\mathcal{I}$ on $V_m$ for the MM1 WT, MM2 WT, MM1 MUT, MM2
MUT, and TP06 models; for the TP06 model $\mathcal{A} \ = m_{\infty}^3$ and
$\mathcal{I} \ = \ j_{\infty} \times h_{\infty}$. 

If we contrast the plots of $\mathcal{A}$ in Figure \ref{fig:actinact}C, we see
that the curves for both MM1 WT and MM1 MUT models lie to the right of, and are
less steep than, their TP06, MM2 WT, and MM2 MUT counterparts; hence,
activation occurs most slowly (with respect to $V_m$) in
MM1 WT and MM1 MUT models. In particular, the Na channels in the TP06, MM2 WT,
and MM2 MUT models reach near-complete activation at $V \simeq \ 2 mV$, $V
\simeq \ -14 mV$, $V \simeq \ -8 mV$, respectively, whereas the MM1 WT and MM1
MUT models do so only for $V \ \simeq \ 60mV$ and $V \ \simeq \ 60mV$,
respectively.  The potential at which $\mathcal{A} \ = \ 0.5$ is
$V_{\mathcal{A}=0.5, \text{TP06}} \simeq -36 mV$, $V_{\mathcal{A}=0.5, \
\text{MM2 WT}} \ \simeq \ -32 \ mV$, $V_{\mathcal{A}=0.5, \ \text{MM2 MUT}} \
\simeq \ -24 \ mV$, $V_{\mathcal{A}=0.5, \text{MM1 WT}} \simeq -20 \ mV$, and
$V_{\mathcal{A}=0.5, \ \text{MM1 MUT}} \ \simeq \ -34 \ mV$.   

From Figure~\ref{fig:actinact}D,  we find that the plots of $\mathcal{I}$, for
the MM1 WT, MM1 MUT, MM2 WT, and MM2 MUT models, are shifted to the right (the
depolarized-potential side) compared to $\mathcal{I}$ for the TP06 model. Note
also that the MM1 MUT model shows inactivation earlier than the MM1 WT model.
In the MM2 MUT case, the inactivation occurs earlier than in the MM2 WT model
in the range $-72 mV  \leq V_m \leq -32 mV$.  The potential at which
$\mathcal{I} \ = \ 0.5$ is $V_{\mathcal{I}=0.5,\text{TP06}} \ \simeq \ -84 \
mV$, $V_{\mathcal{I}=0.5, \text{MM2 WT}} \ \simeq \ -69 / mV$,
$V_{\mathcal{I}=0.5,\text{MM2 MUT}} \ \simeq \ -70 / mV$,
$V_{\mathcal{I}=0.5,\text{MM1 WT}} \ \simeq \ -61 \ mV$ and
$V_{\mathcal{I}=0.5,\text{MM1 MUT}} \ \simeq \ -66 \ mV$.

Our results for the two Markov-state models are in agreement with those shown
in Figure 2 of Refs.~\cite{vecchietti2006computer,carbonell2016comparison}.

\subsection*{Probabilities of the Markov states}

Let us examine now the temporal evolution of the probabilities of different
Markov states during the course of an action potential.  As we have mentioned
in Section~\ref{Methods:MM}, there are three main classes of Markov states for
the Na channel, namely, the open states ($\bf{O,BO}$), the inactivation states
($\bf{IF,IS,IM1,IM2,IC2,IC3}$), and the closed states
($\bf{C1,C2,C3,BC1,BC2,BC3}$). The probabilities of these three classes of
states are as follows: $P_O$ is the open state probability; $P_I$ is the sum
of probabilities of all the inactivation states; and $P_C$ is the sum of
probabilities  of all the closed states. In the case of MM1 WT and MM2 WT
models, they are as follows: 

   $$ Probability =
   \begin{cases}
    P_{O} \ \ \text{(MM1 WT and MM2 WT)}; & \\
    P_{I} = P_{IF} + P_{IM1} + P_{IM2} + P_{IC2} + P_{IC3}  \ \ \text{(MM1 WT)}; & \\
    P_{I} = P_{IF} + P_{IS} + P_{IC2} + P_{IC3}   \ \ \text{(MM2 WT)};  & \\
    P_{C} = P_{C1} + P_{C2} + P_{C3}  \ \ \text{(MM1 WT and MM2 WT)}. & \\
    \end{cases}$$

The probabilities of these three main classes of states, in MM1 MUT and MM2 MUT
models, are as follows:

   $$ Probability =
   \begin{cases}
   P_{O} \ \ \text{(MM1 MUT)}; & \\
    P_{I} = P_{IF} + P_{IM1} + P_{IM2} + P_{IC2} + P_{IC3} \  \ \text{(MM1 MUT)}; & \\
    P_{C} =  P_{C1} + P_{C2} + P_{C3} \ \ \text{(MM1 MUT)}; & \\
    P_{O} \equiv P_{O} + P_{BO} \ \ \text{(MM2 MUT)}; & \\
    P_{I} = P_{IF} + P_{IS} + P_{IC2} + P_{IC3}   \ \ \text{(MM2 MUT)};  & \\
    P_{C} = P_{BC1} + P_{BC2} + P_{BC3} + P_{C1} + P_{C2} + P_{C3} \ \ \text{(MM2 MUT)}.
    \end{cases}$$

\begin{figure}
 \resizebox{\linewidth}{!}{
	\includegraphics[scale=0.65]{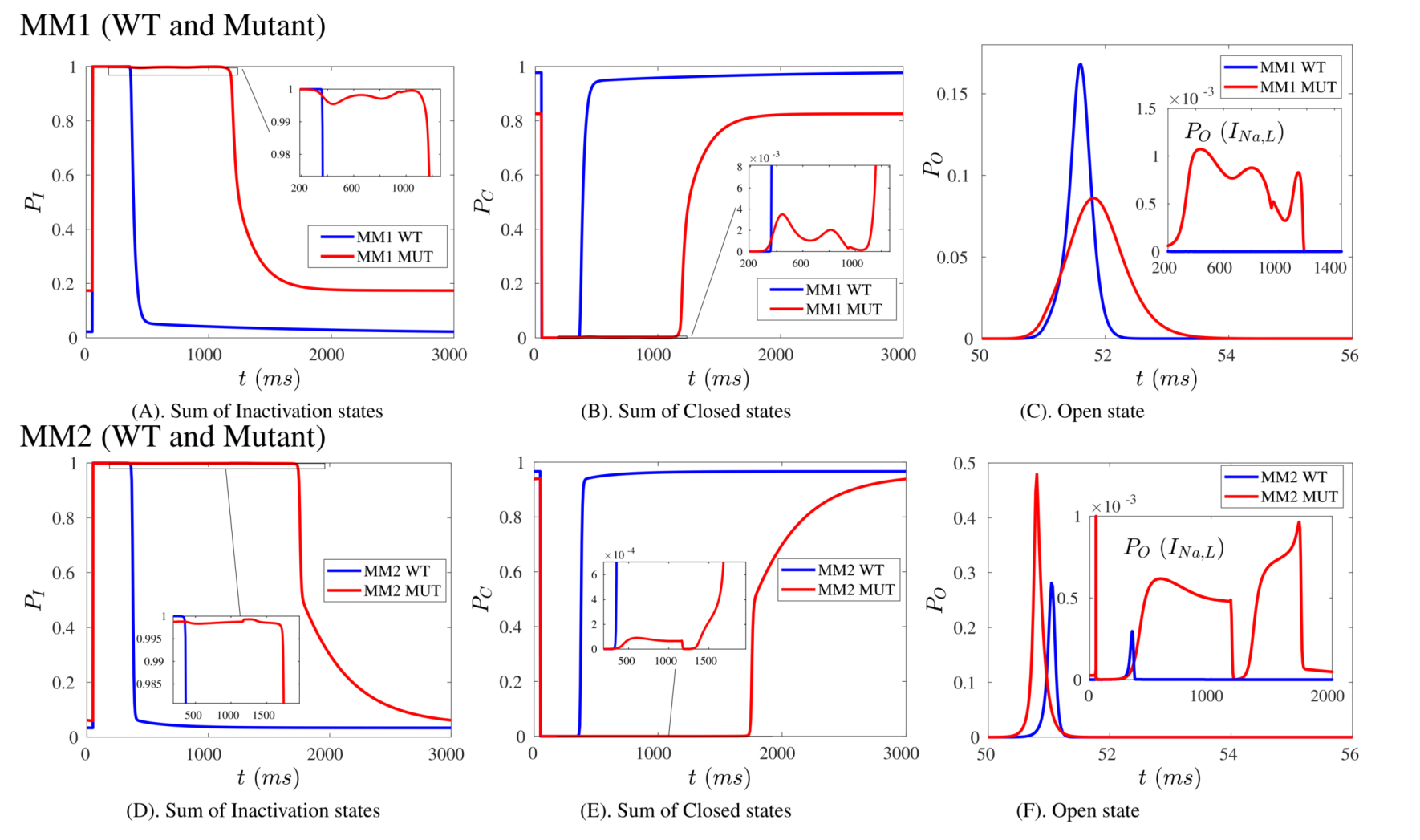}}
    
    \caption{\textbf{Plots of the probabilities $P_{I}$, $P_{C}$, and $P_{O}$
(see text) versus time $t$ for the Na channel in the course of an action
potential.} Plots for the MM1 (MM2) model are in the top (bottom) panel; the
blue and red curves are for WT and MUT models, respectively. We obtain these
plots by pacing a single cell with a pacing-cycle length $\text{PCL} \ = \ 3000 \ ms$
in MM1 and MM2 models for both WT and MUT cases (plots for the $n=501$
stimulation).  The plots in the insets show the sudden opening of the Na
channel in the mutant cases because of delayed inactivation and closing.}

	\label{fig:MMstates}
	\end{figure}

Figure ~\ref{fig:MMstates} shows plots of $P_{O}, \, P_{I},$ and $P_{C}$ versus
time $t$ for the Na channel in the course of an action potential for MM1 (MM2)
models in the top (bottom) panel; the blue and red curves are for WT and MUT
models, respectively. We obtain these plots by pacing a single cell with $PCL \
= \ 3000 \ ms$.  By comparing the blue curves in Figures~\ref{fig:MMstates} (A), (B) , (D) and (E)) we find that the duration for which the Na channel is in the
inactivation or closed states, i.e., the time interval during which $P_I = 1$
and $P_C = 0$ (inactivation state) or $P_I = 0$
and $P_C = 1$ (closed state), is approximately the same in MM1 WT and MM2 WT models. In
contrast, the duration for which $P_O$ is significantly greater than $0$
differs in MM1 WT and MM2 WT models (compare the blue curves in
Figures~\ref{fig:MMstates} (C) and (F)); this duration, measured by the
full-width-at-half-maximum (FWHM) of $P_O$, is $\simeq 0.37 \ ms$ and
$P_{O,max} \simeq 0.17$, for MM1 WT, and $\simeq 0.13 \ ms$ and $P_{O,max}
\simeq 0.28$, for MM2 WT.  The blue curves in the insets of
Figures~\ref{fig:MMstates} (C) and (F) show that, in the MM1 WT model, there is
no late-Na current because $P_O = 0$ in the repolarisation phase of the action
potential (AP); in contrast, the MM2 WT model yields $P_O \simeq 0.0003$ at $t
\ \simeq \ 350 \ ms$, which demonstrates that the Na channel opens in the
repolarisation regime of the AP.

In the MUT cases, the time duration for which the Na channel is in the
inactivation or closed states is prolonged compared to that in the WT cases
(see Figures~\ref{fig:MMstates} (A), (B), (D), and (E)); from the insets of
these figures we see that $P_I$ decreases slightly below $1$ [there are
corresponding increases in $P_O$ and $P_C$ (see Figures~\ref{fig:MMstates} (C)
and (F))], for $ 360 ms \lesssim t \lesssim 1110 ms$ in MM1 MUT and $360 ms \lesssim t
\lesssim 1700 ms$ and MM2 MUT. The duration for which the Na channel is in the
inactivation state $P_I =1$ and $P_C =0$, for MM2 MUT, is much longer than that
in MM1 MUT. We find the following FWHMs: for $P_I$ FWHM $\simeq 1729.2 ms$ 
(MM2 MUT) and $\simeq 1200.25 ms$ (MM1 MUT); for $P_C$ FWHM $\simeq 1721.2 ms$ 
(MM2 MUT) and $\simeq 1176.2 ms$ (MM1 MUT). $P_{O,max}$ is markedly different 
in both MUT models: $\simeq 0.4796$ (MM2 MUT) and $\simeq 0.0861$ (MM1 MUT); 
and the FWHM of $P_O$ is $\simeq 0.14 ms$ (MM2 MUT) and $\simeq 1.03 ms$ 
(MM1 MUT).

\subsection*{Action Potential, APDR, and CVR}

\begin{figure}
 \resizebox{\linewidth}{!}{
	\includegraphics[scale=0.65]{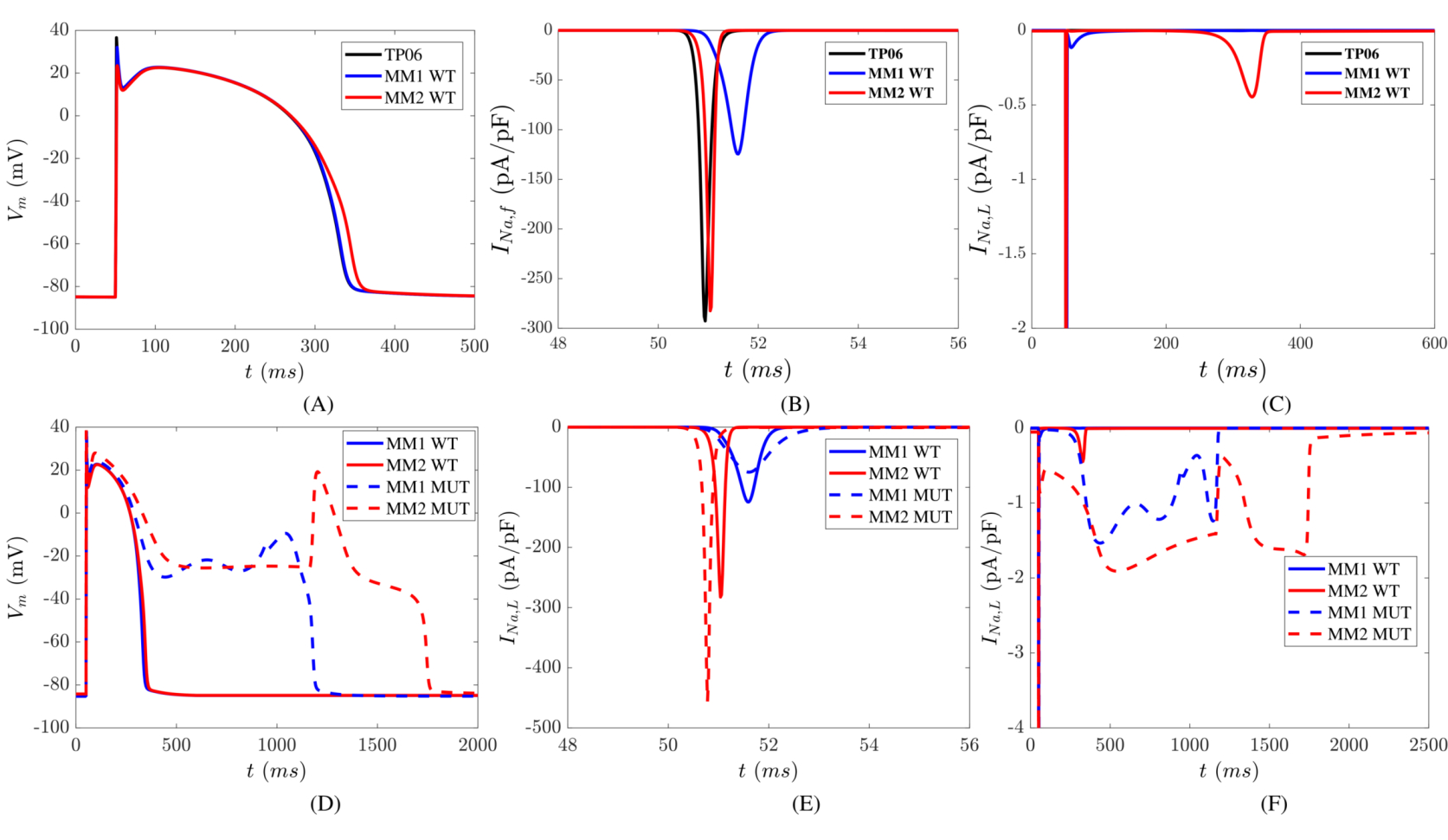}}
	
    \caption{\textbf{Plots of action potentials, the fast Na current $I_{Na,f}$, and
the late Na current $I_{Na,L}$.} Top panel: cell paced for $\text{PCL} =1000 \ ms$ for 
$n=501$ stimulations for TP06, MM1 WT, and MM2 WT models (plots for the $n=501$
stimulation). Bottom panel: cell paced for $\text{PCL} =3000 \ ms$ 
for  MUT models and comparing MM1 WT with MM1 MUT and MM2 WT with MM2 MUT 
(plots for the $n=501$ stimulation). Note that the late opening of the mutant Na 
channel in the repolarization regime causes a release of $I_{Na,L}$ that leads, in
turn, to early afterdepolarizations (EADs) in the AP for mutant models.} 
     
\label{fig:APs}
\end{figure}

We pace a single cell with the following three different values of PCL: high
frequency (PCL=300 ms), intermediate frequency (PCL=650 ms), and low frequency
(PCL=1000 ms).  We present the steady-state AP and the Na current $I_{Na}$ at
the top panel of Figure~\ref{fig:APs}(A),(B), and (C) (for PCL = 1000 ms), for
the three WT models; and we compare the morphological properties of the APs of
these models in Table~\ref{Table2}.

\subsubsection*\underline{{\textbf{PCL $=1000 ms$}}}

Given the differences in the activation profiles in
Figures~\ref{fig:actinact}(C),(D)  and the plots of $P_O$ for the MM models in
Figures~\ref{fig:MMstates}(C).(F) , we observe that (a) the times at which the
Na channels open are different in all the three WT models; and (b) the
amplitude of $I_{Na}$ is comparable in MM2 WT ($ -312.74 \ pA/pF$) and TP06 ($
-300.43 \ pA/pF$) models, but it is significantly lower in the MM1 WT model ($
-144 \ pA/pF$) as we show in Figure~\ref{fig:APs}(B) .  These differences in
the amplitude of $I_{Na}$ affect the maximum voltage and the upstroke-velocity
of the AP (Table~\ref{Table2}). The upstroke velocities for TP06, MM1 WT, and
MM2 WT models are markedly different (Table~\ref{Table2}). Also, there is the late
component of the Na current $I_{Na,L}$ (Figure~\ref{fig:APs}(C)) in the case of
MM2 WT; this component is clearly absent in TP06 and MM1 WT models. $P_O$
becomes significant at $\simeq 350ms$ in the MM2 WT model
(Figure~\ref{fig:MMstates}(F)), so the APD for this model is larger than its
counterparts in the TP06 and MM1 WT models (see Figure~\ref{fig:APs}(C) and
Table~\ref{Table2}).

\subsubsection*\underline{{\textbf{PCL} $=300 ms$}}

As we decrease PCL, say to $300 ms$, we find that both $I_{Na_{max}}$ (the
maximal value of $-I_{Na}$) and the upstroke velocity in  the MM2 WT model
increase relative to their counterparts in the TP06 model as we show in
Table~\ref{Table2} (contrast this with our results for PCL $=1000 ms$). These
increases occur principally because $P_{O,max}$ is higher in the MM2 WT model
than in the TP06 model. 

\subsubsection*\underline{{\textbf{PCL} $= 3000 ms$, MUT Na channel}}

In Figures~\ref{fig:APs}(D),(E) and (F), we show, respectively, plots of $V_m$,
$I_{Na_f}$, and $I_{Na_L}$ versus time $t$; we use dashed curves for the MM1
MUT (blue) and MM2 MUT (red) models and the illustrative value  PCL $=3000 ms$.
Clearly, the MM1 MUT and MM2 MUT APs in  Figure~\ref{fig:APs}(D) show early
afterdepolarizations (EADs)~\cite{zimik2015comparative,vandersickel2014study},
insofar as their APs are prolonged considerably relative to the the APs for MM1
WT and MM2 WT models, because of the failure of inactivation near the
repolarisation region (insets in Figure~\ref{fig:MMstates}). 

\subsubsection*\underline{\textbf{APDR and CVR (WT)}}
\label{apdrcvr}

\begin{figure}
    \resizebox{\linewidth}{!}{
	\includegraphics[scale=0.45]{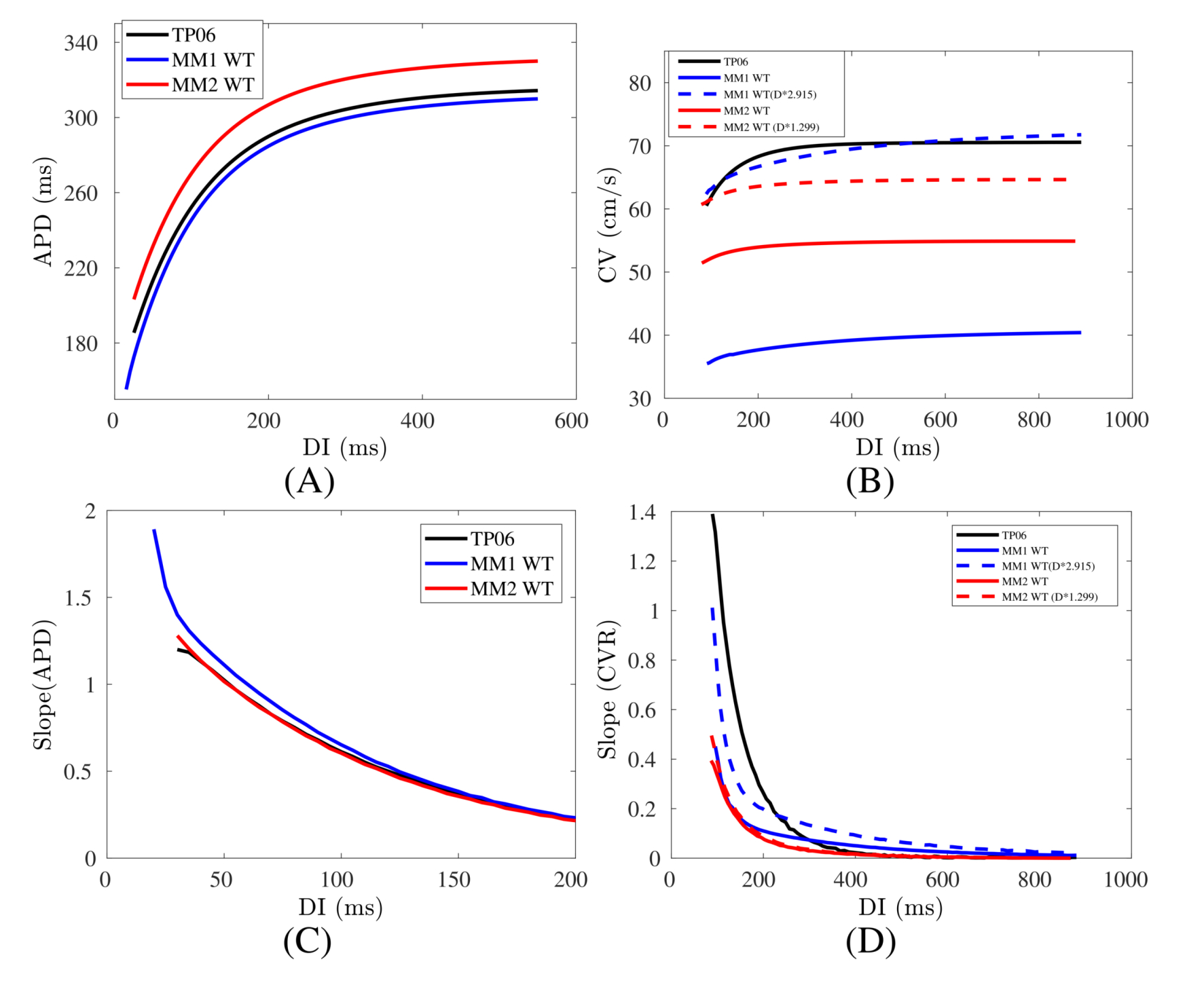}}

	\caption{\textbf{Restitution plots for the TP06, MM1 WT and and MM2 WT
models.} (A) Single-cell static APDR (profiles for TP06 and MM1 WT models lie
close to each other, but the MM2 WT curve lies above these) and (B) the dynamic
CVR, for a one-dimensional cable of cells ($640 \times 10$). The slopes of the
profiles are given in (C) for the APDR and in (D) for the CVR; in all these
three models, the maximal slope of the APDR profile $> \ 1$; the CVR profiles
in MM1 WT and MM2 WT models do not depend sensitively on DI over the range of
values in these plots. The solid lines are calculated for the same diffusion
constant $D$; for the MM1 WT and MM2 WT models this yields steady-state CVs of
$ 40.41 \ cm/s$ and  $54.89 \ cm/s$, respectively, which are not in the normal
range (for the myocardium) $\simeq 60-75 \ cm/s$; if we increase $D$, for the
MM1 WT and MM2 WT models (see text) CV can be brought to this normal range, as
we show by the dashed-line plots.}
	
    \label{fig:restitution}
	\end{figure}

For the TP06, MM1 WT, and MM2 WT models, we present plots of the single-cell
static APDR (Figure~\ref{fig:restitution}(A)) and the dynamic CVR
(Figure~\ref{fig:restitution}(B)), for a one-dimensional cable of cells.  The
APDR profiles for TP06 and MM1 WT lie close to each other, but the MM2 WT curve
lies above these, because of the late current component $I_{Na,L}$ (see above).
The slopes of the APDR and CVR profiles are given, respectively, in
Figures~\ref{fig:restitution}(C) and (D). Note that, in all these three models,
the maximal slope of the APDR profile $> \ 1$ (it is highest in the MM1 WT
model).  The Na channel determines the upstroke velocity at the cellular level;
therefore, this channel plays an important role in determining CV, in cardiac
tissue, and also CVR plots (Figure~\ref{fig:restitution}(B)). From these plots
we find that, for TP06, MM1 WT, and MM2 WT models, CV is nearly independent of
DI, for large DI; the ranges spanned by CV are  $60.51-70.55 \ cm/s$ (TP06),
$35.5- 40.4 \ cm/s $ (MM1 WT), and $51.43-54.89 \ cm/s$ (MM2 WT), for DI in the
interval $90-900 \ ms$; and the saturation values of CV are $\simeq  70.55 \
cm/s $ (TP06), $ \simeq 40.41 \ cm/s$ (MM1 WT), and $ 54.89 \ cm/s$ (MM2 WT).
In the human myocardium, CV is $\simeq 60-75 \ cm/s$ ~\cite{ten2004model,ten2006alternans}.
To obtain CV in this physiological range, we must increase the diffusion
constant $D$ in both MM1 WT and MM2 WT models; we find that, if we multiply $D$
by $2.915$ (MM1 WT) and $1.299$ (MM2 WT), then the saturated value of CV is
$\simeq 64.65 \ cm/s$ (MM1 WT) $\simeq 71.75 \ cm/s$ (MM2 WT); these
multiplicative scale factors can be obtained by noting that $ \text{CV} \propto
\ \sqrt[]{D}$~\cite{kleber2004basic,majumder2011overview} and by using the
saturated CV value in the TP06 model.  With these changes in $D$, CV can be
brought to a physiologically realistic value; but its variation is small:
$60.69-64.65 \ cm/s$ (MM1 WT) and $62.34-71.75 \ cm/s$ (MM2 WT) over the DI range of
$90-900 ms$.

\subsection{2D results}

We have explored differences between the TP06, MM1, and MM2  models at the
single-cell and the cable levels. We now compare spiral- and scroll-wave
dynamics in these  models by carrying out detailed numerical simulations in 2D
(Section ~\ref{sec:2D}) and 3D (Section ~\ref{sec:3D}) domains. 

\label{sec:2D}

      \subsection*{Wild-type Na channel}

\begin{figure*}
\resizebox{\linewidth}{!}{

\includegraphics[width=0.25\linewidth]{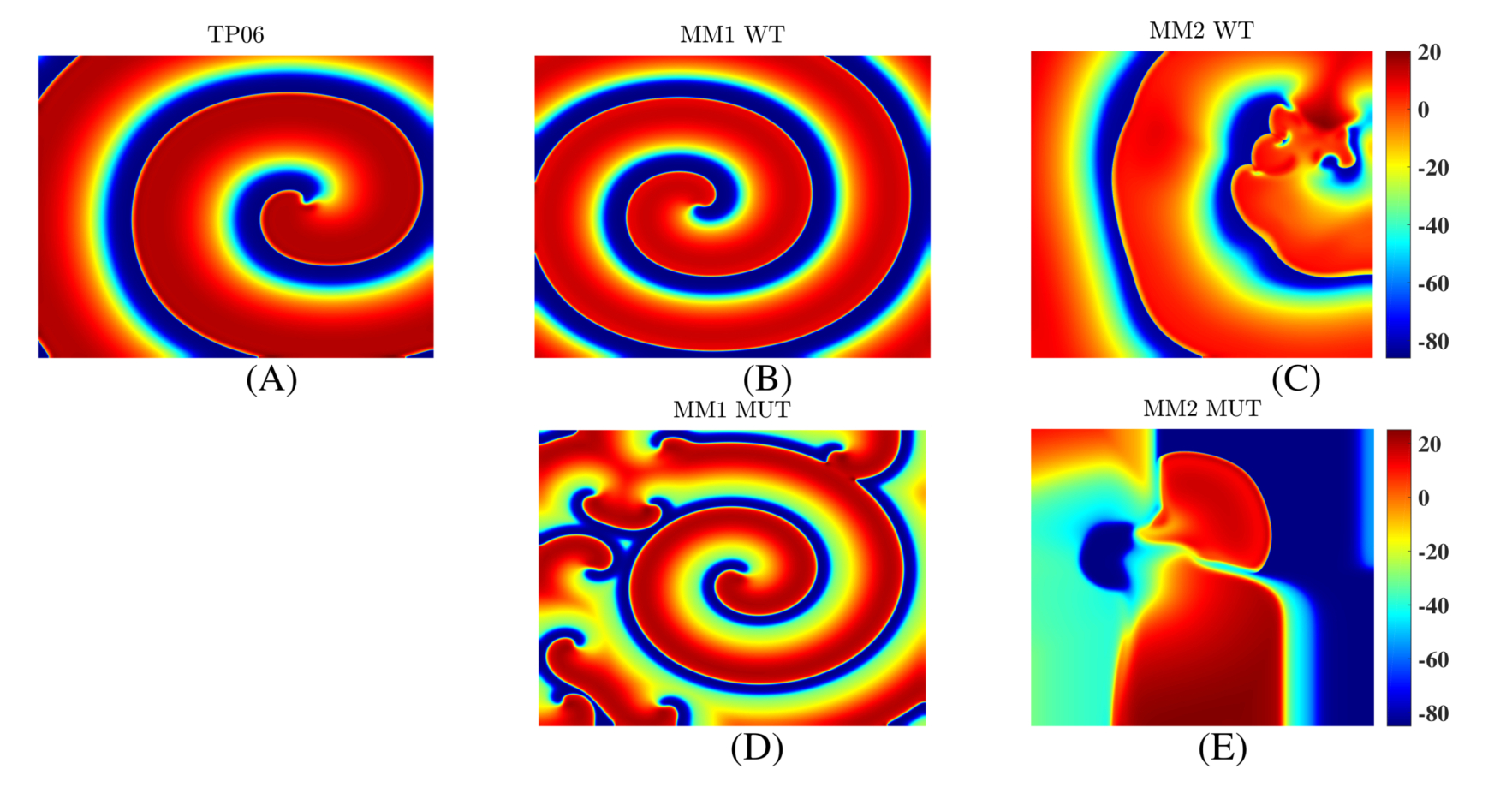}} 
	\caption{\textbf{Pseudocolor plots of the transmembrane potential $V_m$
illustrating the spatiotemporal evolution of spiral waves.} First row: WT models (see text)
(A) TP06, (B) MM1 WT, and (C) MM2 WT; in the MM1 WT (MM2 WT) model 
the spiral wave is stable (unstable). Second row: MUT models (see text) (D) MM1 MUT 
and (E) MM2 MUT. The formation of type-3 EADs (see text) leads to backward propagation 
in the MM2 MUT; by contrast, the type-2 EAD in the MM1 MUT model does not lead
to such backward propagation. However, these EADs create, at the cellular level, dynamical 
heterogeneities, far from the stable spiral core, as shown in (D); this leads to spiral 
break up in a homogeneous MM1 MUT simulation domain. The complete spatiotemporal evolution of the spiral waves of MM1 MUT and MM2 MUT is shown in the \textbf{Movie(M3)} (5
frames per second (fps)) in the Supplementary Material ~\ref{Supp}.}
    \label{fig:spirals}
 \resizebox{\linewidth}{!}{
  \includegraphics[scale=0.5]{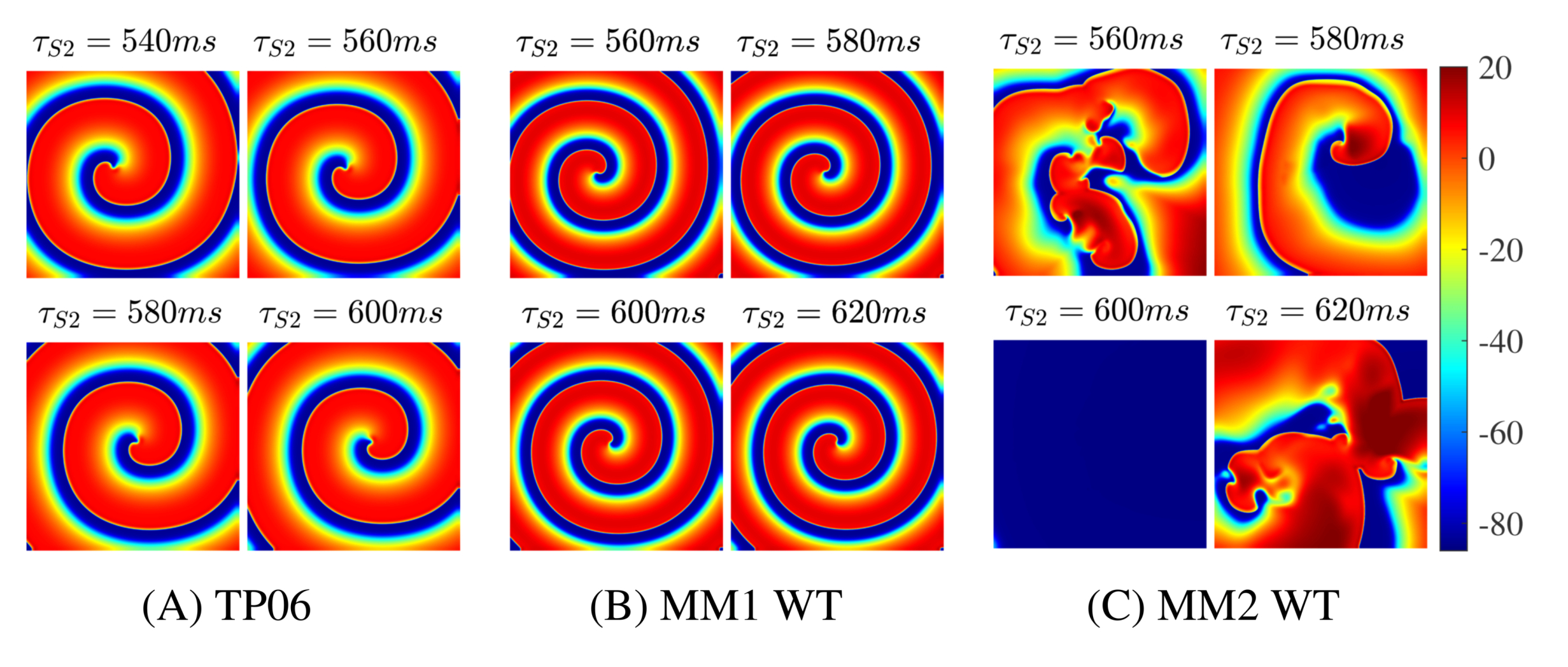}}
	
	\caption{\textbf{Pseudocolor plots of the transmembrane potential
$V_{m}$ illustrating the spatiotemporal evolution of spiral waves for different
values of $\tau_{S2}$, the time interval between the S1 and S2 impulses (see
text).} (A) TP06 and (B) MM1 WT models at $5.9 s$ after spiral-wave initiation;
and (C) the MM2 WT model at $6.9s $ after such initiation. Clearly, the
spatiotemporal evolution of the spiral waves in the TP06 and MM1 WT models is
independent of $\tau_{S2}$, in the range of values investigated here, but not
so for the MM2 WT model. The spatiotemporal evolution of the spiral waves for
different $\tau_{S2}$ of MM2 WT is shown in the \textbf{Movie(M2)} (5 fps) in
the Supplementary Material~\ref{Supp}.}
\label{fig:S2interval}

\end{figure*}

We contrast, in the top panel of Figure~\ref{fig:spirals}, spiral waves in these
three models, with $D \ = \ 0.00154 \ cm^{2}/ms$. We find that spiral waves in
TP06 and MM1 WT are stable and they rotate with frequencies $ \omega \simeq
4.75 \ Hz$  and  $\omega \simeq 4.25 \ Hz$, respectively; in particular, the
low value of CV ($40.41 \ cm/s$), in the MM1 WT model with  $D \ = \ 0.00154 \
cm^{2}/ms$, does not alter the spiral-wave dynamics qualitatively. By contrast,
in the MM2 WT model, the spiral wave is unstable and exhibits transient
breakup; it is not possible to isolate a single cause for this break up, 

but the late Na current $I_{Na,L}$(Figure~\ref{fig:APs}(C)) plays an important
role in this instability; we have checked that, by increasing $\beta_{12}$, we
can reduce the magnitude of this late current and thus suppress spiral-wave
turbulence (the spiral meanders but does not break up into multiple spirals as
we show in the \textbf{Movie (M0)} in the Supplementary Material~\ref{Supp}).

We have carried out another set of studies in 2D simulation domains, with the
values of $D$ scaled up to $D*2.915$ (MM1 WT) and $D*1.299$ (MM2 WT), to bring
the values of CV close to the range of values in human ventricular tissue
~\cite{ten2006alternans}.  These scaled values of $D$ do not change our
qualitative results about spiral-wave stability (TP06 and MM1 WT) or their
breakup (MM2 WT). However, the spiral-arm width increases when we scale up the
value of $D$ (Figure (S1) and \textbf{Movie (M1)} in the Suppelmental
Material~\ref{Supp}); furthermore, because CV increases when we scale up $D$,
the spiral-wave rotation frequency $\omega$ also increases with $D$.
Henceforth, in our 2D and 3D simulations we use the same fixed value $D \ = \
0.00154 \ cm^{2}/ms$ for all three models (TP06, MM1 WT, and MM2 WT).  

We employ the S1-S2 protocol to initiate spiral waves in all these models
(Section~\ref{sec:Methods}). The pseudocolor plots of $V_m$  in
Figure~\ref{fig:S2interval} show that the spiral-wave activity in the TP06 and
MM1 WT models is independent of the time $\tau_{S2}$, at which the S2 pulse is
applied after the S1 pulse (we use $560 ms \leq \tau_{S2} \leq 620 ms$). By
contrast, in the MM2 WT model, we observe spiral-wave breakup for $\tau_{S2} =
560 ms$ and $580 ms$ until the end of our simulation, i.e., $10 s$; but
spiral-wave activity vanishes for $\tau_{S2} = 600 ms$ at $\simeq 5.5s$ and for
$\tau_{S2} = 620ms$ at $\simeq 6.8s$ (Figure~\ref{fig:S2interval} and
\textbf{Movie (M2)} in the Supplementary Material~\ref{Supp}). 

\begin{figure}
    \resizebox{\linewidth}{!}{ 
	\includegraphics[scale=0.6]{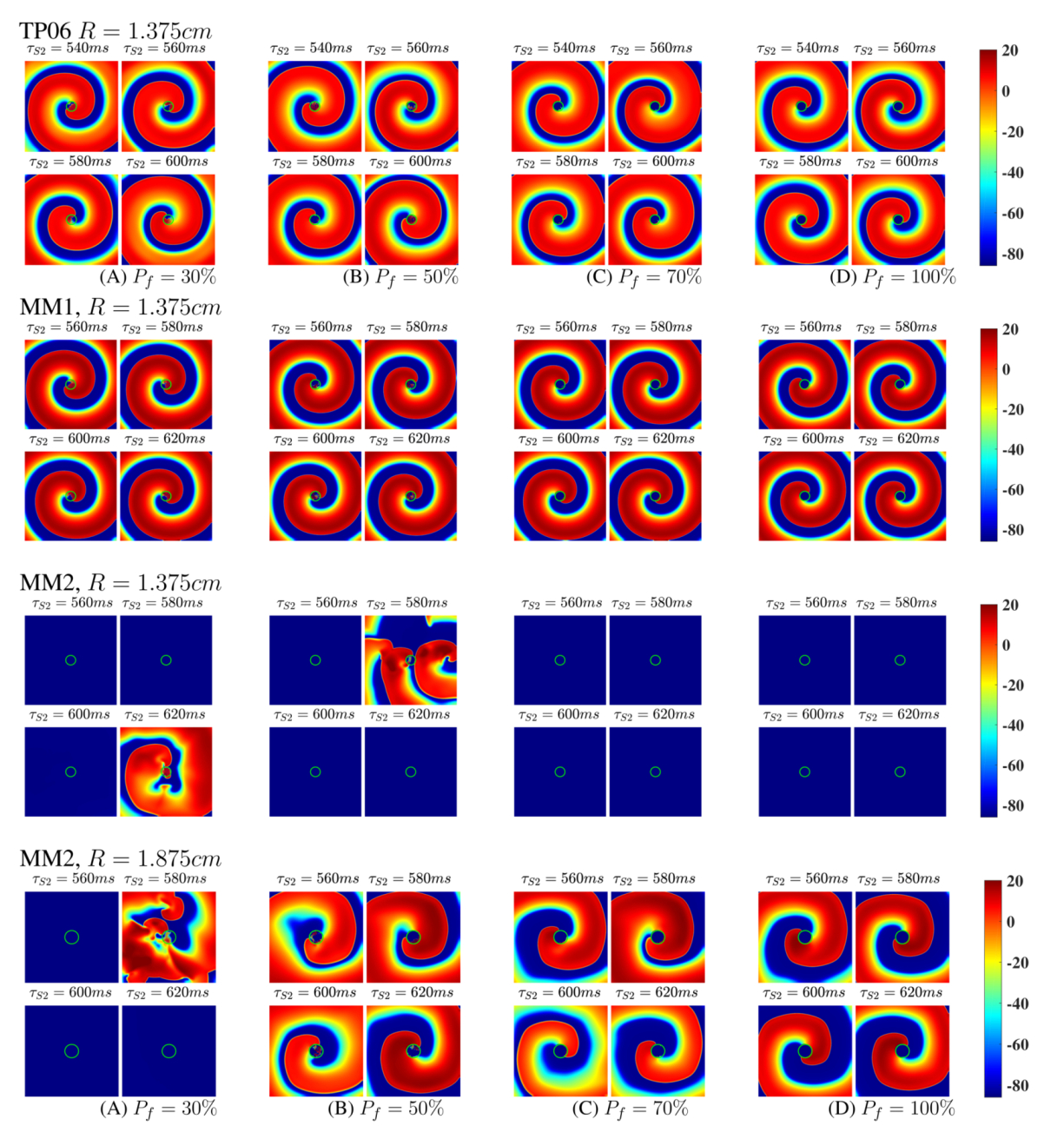}}

	\caption{\textbf{Pseudocolor plots of the transmembrane potential $V_m$
illustrating the spatiotemporal evolution of spiral waves in the presence of
in-excitable obstacles for different values of $P_{f}$ and different
$\tau_{S2}$ (see text) in TP06, MM1 WT and MM2 WT models}. The spiral waves in
TP06 (first row) and MM1 WT (second row) models anchor to the obstacle (for the
values of $P_f$ and $\tau_{S2}$ used here) and so are independent of
$\tau_{S2}$. By contrast, the waves in the MM2 WT model (last two rows) depend
on $R$, $P_f$, and on $\tau_{S2}$. The spatiotemporal evolution of the spiral
waves in the presence of inexcitable obstacles for two representative radius
($R = 1.375$ and $R = 1.875 cm $) for the MM2 MUT model is shown in the
\textbf{Movie(M5,M4)} (5 fps) in the Supplementary Material ~\ref{Supp}. }

	\label{fig:spiralswithobst}
\end{figure}

Spiral-wave dynamics in the MM2 WT model depends on the time $\tau_{S2}$ at
which we initiate the S2 pulse. It behooves us, therefore, to examine whether
obstacles (or conduction inhomogeneities) affect spiral-wave activity in the
MM1 WT and MM2 WT models, for it has been shown, for HH-type models for cardiac
tissue, that spiral-wave dynamics depends sensitively on the position, size,
and shape of such
obstacles~\cite{shajahan2007spiral,shajahan2009spiral,majumder2014turbulent,zimik2017reentry}.
Our obstacles consist of inexcitable points that are distributed randomly
within a circular region of radius $R$; $P_{f}$ is the percentage of the area
of the circle that has inexcitable obstacles. Given our experience with studies
of spiral-wave dynamics with such obstacles in HH-type models, we expect that,
as $P_f$ increases, such an obstacle should anchor a spiral
wave~\cite{lim2006spiral,ikeda1997attachment}.  Therefore, we investigate the
dependence of spiral-wave dynamics on $P_f$ and ${R}$ in the MM1 WT and MM2 WT
models and compare this with its counterpart in the TP06 model, for different
values of $\tau_{S2}$. Illustrative plots from our simulations are shown in
Figure~\ref{fig:spiralswithobst}.  

\begin{figure}  
    \resizebox{\linewidth}{!}{ 
	\includegraphics[scale=0.99]{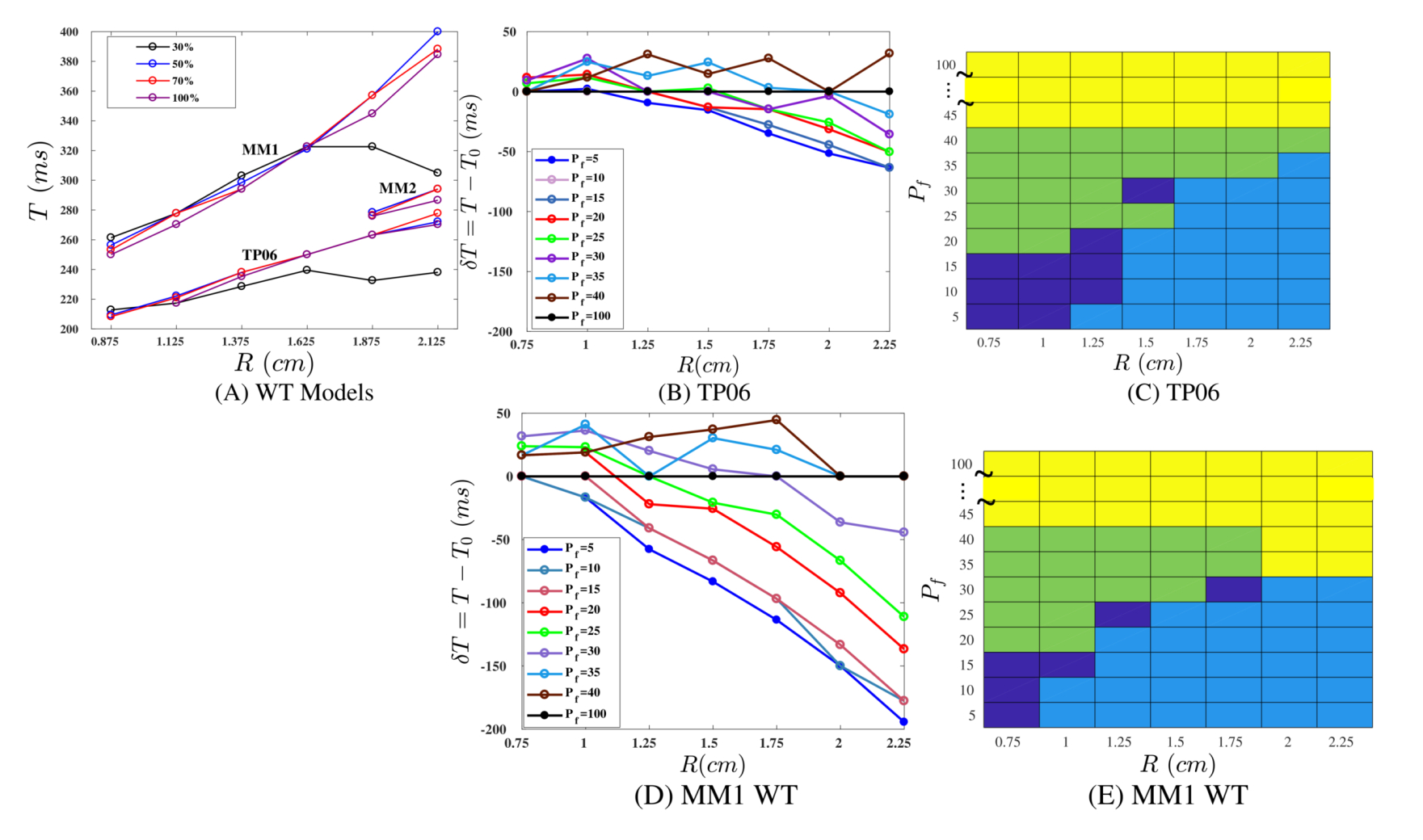}	
} \newline

	\caption{\textbf{The dependence of the time period $T$ of the anchored
spiral wave on the radius $R$ and percentage of fibrosis $P_f$ .} (A) Plots of
$T$ versus $R$ for different values of $P_f$ for TP06, MM1 WT and MM2 WT
models; note spiral anchoring starts around $R \simeq 1.875 cm$ for the MM2 WT
model. (B) and (D): Plots of the change in time period $\Delta T$ versus,
(obtained from five recording points in the domain of which one grid point is
in the region with heterogeneity) for TP06 and MM1 WT models ($T_0$ is the time period
for a completely inexcitable obstacle ($P_f = 100 \%$), for different $P_f$. If
$\Delta T > 0$, then the frequency $\omega \sim T^{-1}$, for a given pair
($R,P_f$), is less than $\omega_0 \sim T_0^{-1}$ (for $R, P_f =100\%$); this
may occur because the spiral core penetrates the obstacle because of a spanning
cluster of excitable regions inside the obstacle.  
In (C) and (E) we show different regions in the ($R,P_f$) plane for TP06 and
MM1 WT models, respectively, with the following the color code: \textbf{Light
Blue}: an increase in $\omega$ relative to $\omega_0$ (caused by penetration of
the spiral core).  \textbf{Light Green}: decrease in $\omega$ relative to
$\omega_0$  (accompanied by penetration of the spiral core); ($R,P_f$).
\textbf{Yellow}: No penetration of the spiral core into the obstacle.
\textbf{Dark Blue}: No change in $\omega$ relative to $\omega_0$ even though
the spiral core penetrates into the obstacle.}

	\label{fig:radiusvsT}
   
\vspace{0.5cm}
     
    \resizebox{\linewidth}{!}{ 
	\includegraphics[width=0.5\linewidth]{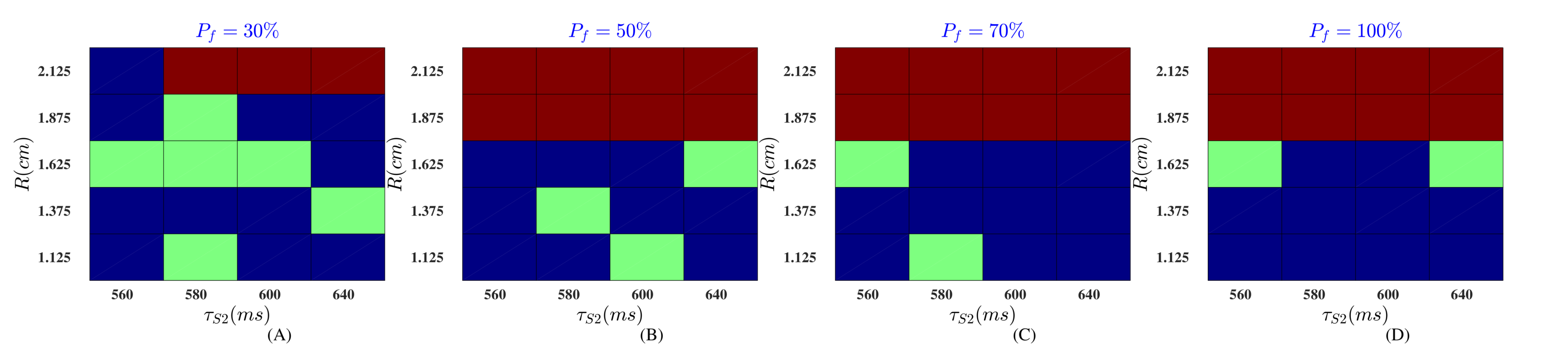}}
    
	\caption{\textbf{Stability diagrams for spiral-wave activity, in the
presence of localized, in-excitable obstacles distributed within a circular
region of radius $R$, for different values of $P_f$, in the MM2 WT model.}
Color code: \textbf{Brown, Green}, and \textbf{Blue} show regions with an
anchored spiral, spiral breakup, and no activity, respectively.}

	\label{fig:phaseplot}
	\end{figure}

We find that, for the TP06 and MM1 WT models, the anchoring of the spiral wave
depends on $R$ and on $P_{f}$, but not on $\tau_{S2}$. The time period $T$ of
the anchored spiral increases with $R$ and $P_{f}$ as we show in
Figure~\ref{fig:radiusvsT}(A); but $T$ decreases for lower percentages (e.g.,
$P_f =30 \%$) in TP06 and MM1 WT models at large values of $R$
(Figure~\ref{fig:radiusvsT}(A)). The interaction of the tip of the spiral wave
with the obstacle is complicated. In particular, this depends on how much of
the region, inside the circular patch, is excitable. For low values of $P_f$,
this excitable region forms a tortuous but spanning cluster (in the sense of
percolation theory~\cite{stauffer1994introduction}), so the tip of the spiral
propagates inside the obstacle, the wave of activation is slightly deformed
there, but then it re-emerges into the homogeneous part of the simulation
domain.  If $P_f$ is large, the excitable region can still be tortuous, but it
does not form a spanning cluster, so the tip of the spiral rotates around the
obstacle, and is anchored to it, but does not propagates inside it.  To
quantify the effect of our obstacle on the spiral wave we calculate $\delta T
\equiv (T-T_0)$, where $T$ is the time period (or inverse of the rotation
frequency $\omega$), at a given set of values of $P_f$ and $R$, $T_0$ is the
time period (or inverse of the corresponding frequency $\omega_0$) with $P_f =
100\%$ for the same value of $R$. Clearly $T$ must depend on $P_f$ and $R$. The
plots in Figures~\ref{fig:radiusvsT}(B)and (D) show, for TP06 and MM1 WT
models, the dependence of $\delta T$ on $R$ for different values of $P_f$.
Given these plots, we identify  three regions, namely, (i) $\delta T < 0$,
i.e., $\omega > \omega_0$, (ii) $\delta T > 0$, i.e., $\omega < \omega_0$ and
(iii) $\delta T = 0$ i.e., $\omega = \omega_0$.  If $\delta T > 0$, then the
frequency $\omega \sim T^{-1}$, for a given pair ($R,P_f$), is less than
$\omega_0 \sim T_0^{-1}$ (for $R, P_f =100\%$); this may occur because the
spiral core penetrates the obstacle because of a spanning cluster of excitable
regions inside the obstacle.  In Figures~\ref{fig:radiusvsT}(C) and (E) we show
different colored regions in the ($R,P_f$) plane for TP06 and MM1 WT models,
respectively: light blue indicates an increase in $\omega$ relative to
$\omega_0$ (caused by penetration of the spiral core);  light green is for a
decrease in $\omega$ relative to $\omega_0$  (accompanied by penetration of the
spiral core); yellow indicates no penetration of the spiral core
into the obstacle; dark blue depicts regions in which there is no change in
$\omega$ relative to $\omega_0$ even though the spiral core penetrates into the
obstacle.

For the MM2 WT model, the minimum size $R_{min}$ for spiral anchoring is large,
compared to that in TP06 and MM1 WT model, and is $R_{min} \ = \ 1.875 \ cm$
(\textbf{Movie (M5)} in the Supplementary Material~\ref{Supp}).  The threshold
percentage in the MM2 WT case is  $P_{f,min} \simeq 50 \%$.  Once we reach the values $R_{min}$
and $P_{f,min}$ required for anchoring, the spiral activity is independent of
$\tau_{S2}$, as we show in the fourth row of Figure~\ref{fig:spiralswithobst}.
The dependence of the spiral rotation time period $T$ on $R$, for different
values of $P_f$, is shown in Figure~\ref{fig:radiusvsT}(A).  Stability diagrams
for the spiral-wave activity, in the presence of localized, inexcitable
obstacles distributed within a circular region of radius $R$,  are shown in the
$(R,\tau_{S2}$) plane, for different values of $P_f$ in the MM2 WT model, in
Figure~\ref{fig:phaseplot}; brown, green, and blue denote regions with an
anchored spiral, spiral breakup, and no activity, respectively.

 \subsection*{Mutant Na channel}

The mutant Na channel fails to inactivate completely in the MM1 MUT and MM2 MUT
models; this leads to prolonged EADs, as we have shown in
Sec~\ref{sec:actinact} and Figure~\ref{fig:APs}. We find that two of the types
of EADs that have been discussed in Ref.~\cite{zimik2015comparative} occur in
both these MUT models: there is a single EAD (of type $2$ in the nomenclature
of Ref.~\cite{zimik2015comparative}), in the MM1 MUT model, and an oscillatory
EAD (roughly of type $3$ in the nomenclature of
Ref.~\cite{zimik2015comparative}), in the MM2 MUT model. These two types of
EADs affect the wave dynamics differently, as we demonstrate explicitly by
simulating plane-wave propagation in our 2D domain, but with all mutant
myocytes. We observe backward propagation of the plane wave in the MM2 MUT
model because of the oscillatory EADs; by contrast, there is no such backward
propagation in the MM1 MUT model. If we initiate a spiral wave in both these
models, then, (a) in the MM1 MUT model, we get almost-instantaneous far-field
breakup away from the core, but the mother rotor is unaffected, and (b) in the
MM2 MUT model, we obtain almost-instantaneous spiral break-up (bottom panels of
Figures~\ref{fig:spirals}(D) and (E)). The complete spatiotemporal evolution of
such spiral-wave dynamics, for both these cases, is shown in the $\textbf{Movie
(M3)}$ in the Supplementary Material~\ref{Supp}. 

\begin{figure*}
	\hspace{1.5cm}
	\includegraphics[scale=0.25]{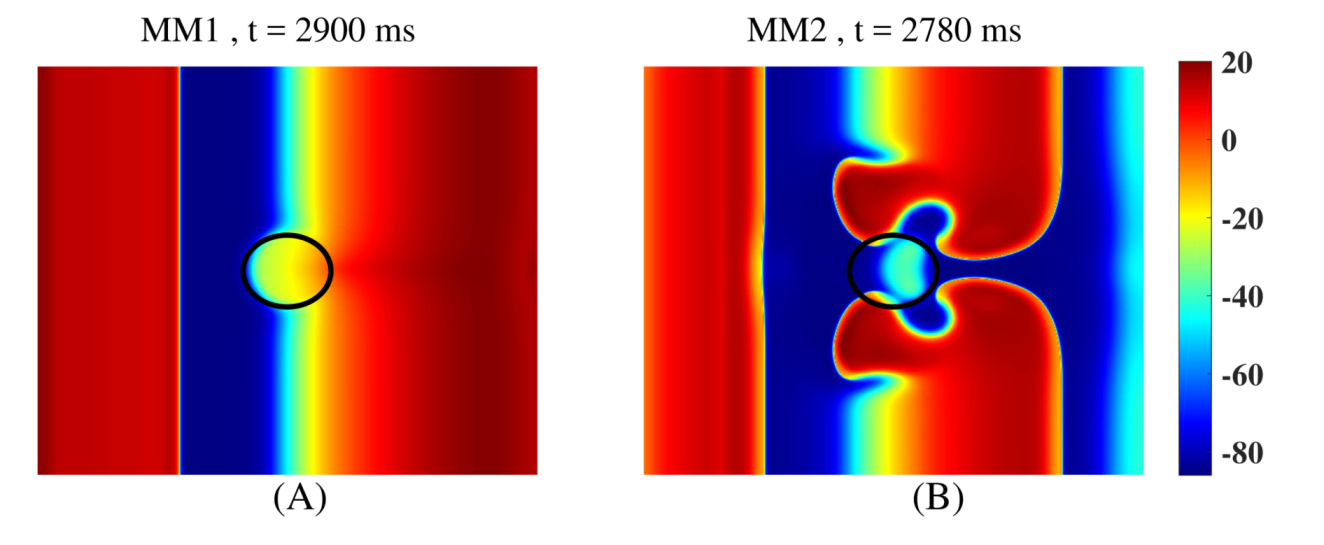}
    
	\caption{\textbf{Pseudocolor plots of the transmembrane potential $V_m$
illustrating the spatiotemporal evolution of electrical-activation waves when there is a
circular clump of mutant cells.} The clump radius $\text{R} = \ 1.125 \ cm$
(shown via a black circle); this clump is surrounded by wild-type (WT) cells,
and the simulation domain is paced from the left boundary (pacing frequency
$3.7 Hz$).  (A) MM1 MUT model (no spiral wave forms); and  (B) MM2 MUT model (a
spiral wave forms).  For the complete spatiotemporal evolution movie see  the
\textbf{Movie(M10)} (5 fps) in the Supplementary
Material~\ref{Supp}. The qualitative difference between (A) and (B) arises
because of  the different types of EADs in MM1 MUT and MM2 MUT models.} 

     \label{fig:MUTpacing}
\end{figure*}

Although there are several studies of the effects of different types of
inhomogeneities on spiral-wave dynamics in mathematical models for cardiac
tissue (see, e.g.,
Refs.~\cite{shajahan2007spiral,shajahan2009spiral,majumder2014turbulent,zimik2017reentry}
and references therein), to the best of our knowledge there has been no study,
based on Markov-state models, of an inhomogeneity comprising mutant myocytes in
a background of wild-type myocytes. Therefore, we present a representative
study of spiral-wave dynamics in the presence of a clump of only mutant cells,
of radius $\text{R} = 1.125 \ cm$, embedded in a background of wild-type cells.
We then explore the possibility of spiral-wave formation via high-frequency
stimulation by pacing the simulation domain from the left boundary (pacing
frequency $3.7 Hz$). We find that, in the MM1 MUT model, no spiral wave forms
(Figure~\ref{fig:MUTpacing}(A)); by contrast, in the MM2 MUT model a spiral wave
forms  (Figure~\ref{fig:MUTpacing}(B)). [The spatiotemporal evolution of these
waves is shown in the \textbf{Movie(M10)} in the Supplementary
Material~\ref{Supp}.] This qualitative difference arises because of the
different types of EADs (discussed above) in MM1 MUT and MM2 MUT models.

   \subsection{3D results} 
\label{sec:3D}

\begin{figure*}

	\includegraphics[scale=0.25]{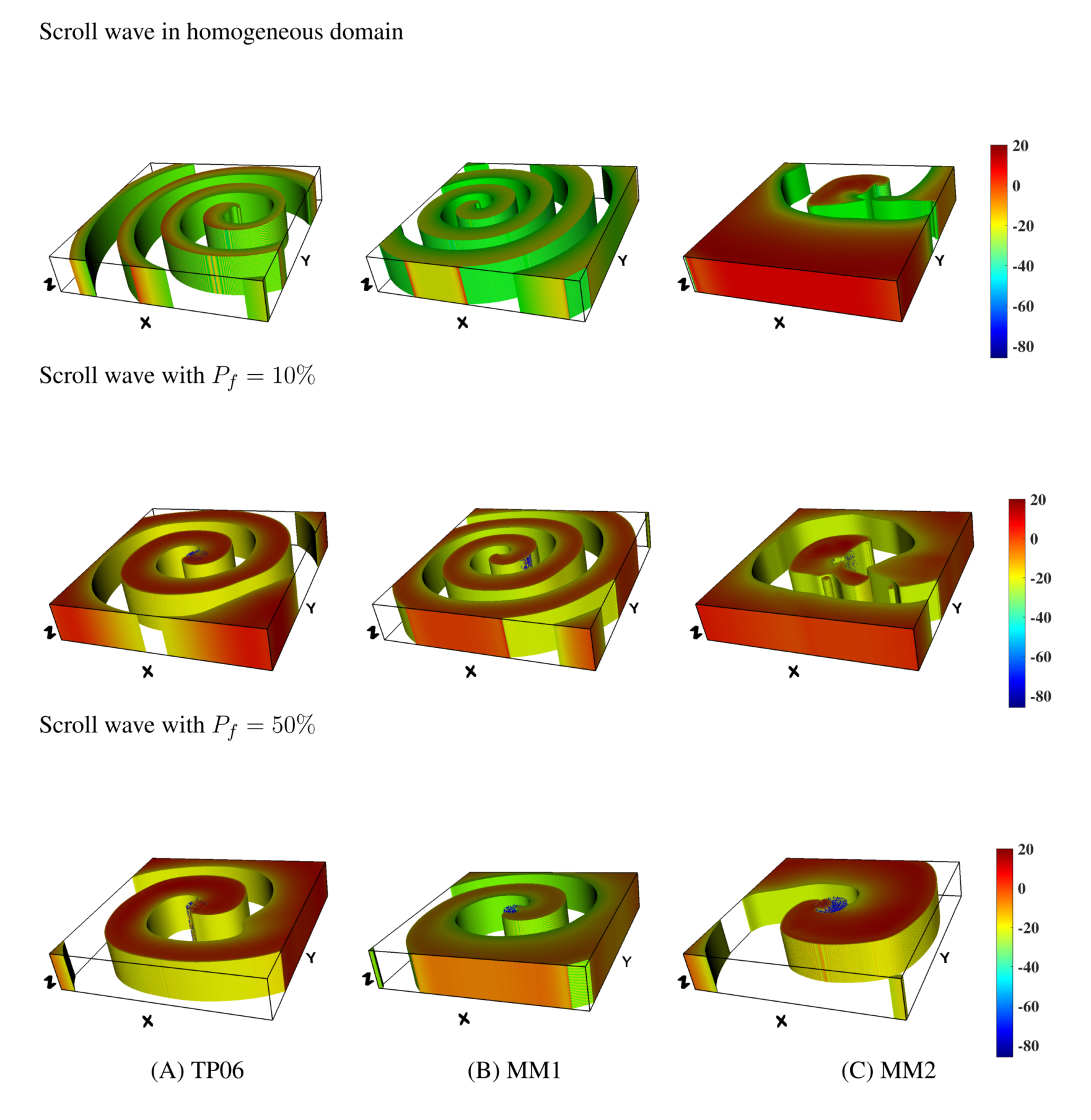}

    \caption{ \textbf{Color isosurface plots of the transmembrane potential
$V_m$ illustrating scroll waves in  (A) TP06, (B) MM1 WT, and (C) MM2 WT
models.} The isosurfaces lie between $-10 mV$ and $30 mV$  in both the
homogeneous domain (top panel) and with localized obstacles [$P_{f} = 10 \% $
(middle panel) and $P_{f} = 50 \% $ (bottom panel)]; these illustrative plots
are at $t   = 2s$.  In a homogeneous domain scroll waves are stable in TP06 and
MM1 WT models, but not in the MM2 model. Note that the scroll wave is anchored
to the obstacle with $P_{f} = 50 \% $ in the MM2 model. For the spatiotemporal
evolution of these scroll waves see \textbf{Movies(M6-M8)} (5 fps) 
in the Supplementary Material~\ref{Supp}.}

	    \label{fig:scrolls}
	\end{figure*}

We end with an illustrative study of scroll waves in 3D TP06, MM1 WT, and MM2 WT
models.  These waves are shown via color isosurface plots of the transmembrane
potential $V_m$ in Figures~\ref{fig:scrolls} A, B, and C, respectively, for
both a homogeneous domain (top panel) and with localized obstacles [$P_{f} = 10
\% $ (middle panel) and $P_{f} = 50 \% $ (bottom panel)].  In a homogeneous
domain, scroll waves are stable in TP06 and MM1 WT models, but not in the MM2
model.  An inexcitable obstacle, with $P_f=10\%$, has no significant impact on
scroll waves in TP06 and MM1 WT models. An increase in $P_f$, say to
$P_f=50\%$, leads to an anchoring of the scroll waves at the obstacle (as in
the study of inexcitable obstacles in Ref.~\cite{majumder2011scroll}). For the
MM2 model, with $P_f = 10 \%$, scroll-wave break-up is enhanced; but for
$P_f=50\%$, the scroll wave gets anchored to the obstacle. The spatiotemporal
evolution of these scroll waves is shown in the \textbf{Movie(M10)} in the
Supplementary Material~\ref{Supp}.

\section{Discussion and Conclusions}

Earlier studies of Markov models for cardiac myocytes have focused on the
effects of mutations in subunits of $Na^+,\ \text{and} \ K^+$ channels
in the context of the Brugada and LQT
syndromes~\cite{irvine1999cardiac,clancy1999linking,clancy2001cellular,clancy2002na+};
these studies have elucidated the effects of changes in the kinetic
properties of these ion-channel, and their consequences, such as the
prolongation of the APD, which leads, in turn, to EADs. Also, Markov
models have been used to asses the importance of a particular
functionality of an ion channel, e.g., the role of $I_{Ks}$ on AP
repolarisation~\cite{silva2005subunit}. Furthermore, Markov models have
been used to investigate the activation and inactivation properties of
$Na^+, \ K^+ \ \text{and} \ Ca^{+2}$ ion channels as, e.g., in
~\cite{wang1997quantitative,bondarenko2004model,wang2005time,bondarenko2004computer}.
In addition, some studies have focused on theraupetics and
drug-channel interactions~\cite{clancy2007pharmacogenetics,moreno2011computational,moreno2013ranolazine}, from cellular to the anatomically realistic tissue levels.

We have investigated, from cellular to tissue levels, the differences in
kinetic properties of Na channels in TP06, MM1 (WT and MUT), 
and MM2 (WT and MUT) models. We have shown that Na channels in TP06
and MM2 (WT and MUT) models are activated faster, with respect to $V_m$, than
their counterparts in the MM1 (WT and MUT) models; also the
inactivation of these channels is faster in the TP06 model than in MM1 (WT and MUT) 
and MM2 (WT and MUT) models. These differences leads to different times 
of openings of the Na channel and the amplitudes of $I_{Na,f}$ are completeley 
determined by the amplitude of $P_O$ in the MM models. These changes in the
amplitudes of $I_{Na,f}$, $I_{Na,L}$, and the activation-inactivation
dynamics lead to disparate CVR and maximal CVs in cable
simulations.  To the best of our knowledge, our study is the first to compare 
spiral-wave dynamics in different Markov models for the Na (WT and MUT)
ion channels in realistic mathematical models for cardiac tissue. 
We have carried out \textit{in silico} studies, in both
homogeneous simulation domains and domains with inhomogeneities, to
compare and contrast spiral- and scroll-wave dynamics in five different
models for cardiac tissue (Hodgkin-Huxley type, TP06
model~\cite{ten2006alternans}, and  Markov-state models such as MM1 WT
and MM2 WT, for the WT Na channel, and MM1 MUT and MM2 MUT, for the
mutant Na
channel~\cite{vecchietti2006computer,moreno2011computational}).  Our
study explores the sensitive dependence of spiral- and scroll-wave
dynamics on these five models and the parameters that define them. We
also examine the control of spiral-wave turbulence in these models. To
the best of our knowledge, such a comparative study of wave dynamics in
HHM and Markov-state models has not been carried out hitherto. In our
opinion, such a comparison is even more valuable than the comparison of
single-cell properties of models for cardiac myocytes.  We hope our
study will lead to more comparisons of wave dynamics in different
mathematical models for cardiac tissue and in \textit{in vitro}
experiments. Furthermore, we have carried out a detailed parameter-sensitivity study, principally for the WT models by using multivariable linear regression (see the Supplementary material).

We mention some of the limitations of our study. We have studied the
differences in Na ion-channel modeling, which is important in the
context of the LQT syndrome; but we have not carried out such a study
for Kr-channel modeling, as mutations in the Kr channel also leads to
the LQT syndrome. We have considered a few, illustrative Markov-state
models for Na channels, given our computational resources; many more
such Markov-state models have been developed for the Na
channel~\cite{fink2009markov,balbi2017single}; a comprehensive
comparison of all these models lies beyond the scope of this paper.
Also, we have used a single base model, i.e., TP06 model upon which we
build the Markov-state models (MM1 WT, MM1 MUT, MM2 WT and MM2 MUT); other base models, e.g., the O'Hara Rudy model~\cite{o2011simulation}, can
be used; a comprehensive comparison of all these models lies beyond the
scope of this paper.  We do not use an anatomically realistic
simulation domain~\cite{trayanova2009integrative}, with information
about the orientation of muscle 
fibers~\cite{majumder2011scroll,majumder2012nonequilibrium}; and we use a
monodomain description for cardiac tissue. These considerations lie
beyond the scope of this paper. We note, though, that the study of
Ref.~\cite{potse2006comparison,bourgault2010comparing} has compared
results from monodomain and bidomain models and has shown that the
differences between them are small.

\section*{Acknowledgments}
{We thank the Department of Science and Technology (DST), India, for the funding and the Supercomputer
Education and Research Centre (SERC, IISc) for computational resources.}

\section*{Author Contributions}

Conceived and designed the study: MKM ARN RP. Performed the simulations: MKM.
Analyzed the data: MKM ARN RP.  Contributed analysis: MKM ARN RP. Wrote the
paper: MKM ARN RP.

\section*{Supplementary Data}

\textbf{Figure.S1}: Spiral waves in the MM1 WT and MM2 WT models, with increased
values of  $D$ (see text). Apart from an increase in the conduction-velocity CV and the
arm-length, the qualitative features of spiral-wave dynamics here remain the same as
those with the unaltered value of $D$. See the \textbf{Movie(M1)}.
\newline

\textbf{Figure.S2}: Pseudocolor plots of the transmembrane potential $V_m$
illustrating the elimination of spiral waves by electrical stimulation on a
square mesh. The domain is divided into square cells of dimension $128\times
128$ grid points (for a domain size $1024 \times 1024$) or $64 \times 64$ (for a domain size $512 \times 512$); and then a stimulus, of amplitude $50 pA/ pF$, is applied
for $100 ms$ along the edges of the square cells as in
Ref.~\cite{sinha2001defibrillation}; this leads to the elimination of
spiral-wave activity in all the five models. 
\newline

\textbf{Movie(M0)}: This movie shows the quasi-stable behavior of spiral waves in
the MM2 WT model when we increase one of the rate constant, namely, $\beta_{12} \to
\beta_{12} *25 $ .

\textbf{Movie(M1)}: This movie (5 frames per second (fps)) shows pseudocolor
plots of $V_m$ that illustrate spiral waves in the MM1 WT and MM2 WT models,
with increased values of  $D$ (see text): (a) MM1 WT(D), (b) MM1 WT
($D$*2.915), (c) MM2 WT ($D$), and (d) MM2 WT ($D$*1.299).  Apart from an
increase in the conduction-velocity CV and the arm-length, the qualitative
features of spiral-wave dynamics here remain the same as those with the
unaltered value of $D$.
\newline

\textbf{Movie(M2)}: This movie (5 fps) shows pseudocolor plots of $V_m$ that illustrate
the sensitive dependence of wave activity
on $\tau_{S2}$ (see text) in the MM2 WT model: (a) $\tau_{S2}  = 560 ms$, (b)
$\tau_{S2} = 580 ms$, (c) $\tau_{S2} = 600ms$, and (d) $\tau_{S2} = 620ms$.  
\newline

\textbf{Movie(M3)}: This movie (5 fps) shows pseudocolor plots of $V_m$ that illustrate
the spatiotemporal evolution of the spiral waves of MM1 MUT and MM2 MUT models.
\newline

\textbf{Movie(M4,M5)}: This movie (5 fps) shows pseudocolor plots of $V_m$  that illustrate the
dependence on $\tau_{S2}$ of spiral-wave initiation, in the MM2 WT model in the
presence of inhomogeneities (see text) with  $P_f = 30,\, 50,\, 70,$ and $100$,
within circular regions of radii $R = 1.875 cm$ and $R = 1.375 cm$ .  \newline

\textbf{Movie(M6-M8)}: These movies (5 fps) show isosurface plots of $V_m$ that
illustrate scroll-waves in homogeneous media in TP06, MM1 WT, and MM2 WT
models.
\newline

\textbf{Movie(M10)}: These movies (5 fps) show isosurface plots of $V_m$ that
illustrate the pacing of the tissue in MM1 and MM2
models, with a circular heterogeneities of Na mutant cells (radius of $R=1.125
cm$) surrounded by the Na wild-type cells.
\newline

\textbf{Movie(M11)}: These movies (5 fps) show isosurface plots of $V_m$ that
illustrate the successful elimination of the spiral waves and spiral-wave turbulence
(by using the method described in the Supplementary material)  for the MM1 and MM2 (WT and MUT) models. 
\newline

 \label{Supp}

\bibliography{references}
\bibliographystyle{unsrt}
 \par

 \subsection{Figures}      


\subsection{Tables}  
   
\begin{table}
\begin{center}
    \begin{tabular}{ | l |  p{8.5cm} |}
    \hline 
    $I_{Na}$ & fast inward $Na^+$ current  \\ \hline 
    $I_{CaL}$ & L-type inward $Ca^{++}$ current   \\ \hline
    $I_{to}$ & Transient outward current   \\ \hline
    $I_{Ks}$ & Slow delayed rectifier outward $K^+$ current    \\ \hline
    $I_{Kr}$ & Rapid delayed rectifier outward $K^+$ current   \\ \hline 
    $I_{K1}$ & Inward rectifier outward $K^+$ current   \\ \hline
    $I_{NaCa}$ & $Na^+ / Ca^{++}$ exchanger  current  \\ \hline
    $I_{NaK}$ & $Na^+ /K^+$ pump current    \\ \hline
    $I_{pCa}$ & plateau $Ca^{++}$ current  \\ \hline 
    $I_{pK}$ & plateau $K^+$ current   \\ \hline
    $I_{bNa}$ & background inward $Na^+$ current   \\ \hline
    $I_{bCa}$ & background inward $Ca^+$ current    \\ \hline
    
    \end{tabular}

 \hspace{7cm}\caption{\textbf{The various ionic currents in the TP06 model Ref.~\cite{ten2006alternans}; the symbols used for the currents follow Ref.~\cite{ten2006alternans}. }}
\end{center}
\label{Table1}
\end{table}     
 
\begin{table}
\begin{tcolorbox}[tab2,tabularx={X|Y|Y|Y|Y|Y|}]
 
$\bf{PCL} $  & $\bf{Model} \ $  & $\bf{INa_{max}} $ & $\bf{V_{max}}  \hspace{0.2cm} \ $ &   $\bf{\frac{dV}{dt}_{max}} \ $ &   $\bf{APD} $  \\    
$\bf{(ms)}$  &   & $\bf{(pA/pF)}$ &  $\bf{(mV)}$ &  $\bf{(mV/ms)}$ &  $ \bf{(ms)}$  \\\hline\hline
   & TP06  & $-177.19 $ & $23.42$  & $227.97$ &   $218.76$ \\\hline\hline
$\bf{300}$ & MM1 WT & $-82.23 $ & $21.17$ &  $81.76$ &   $219.66$ \\\hline\hline
   & MM2 WT & $ -252.13$ & $18.56$ &  $256.8$ &   $226.76$ \\\hline 
     \\\hline\hline
   & TP06 & $ -298.19$ & $37.17$ &  $349.94$ &   $291.56$ \\\hline\hline
$\bf{650}$ & MM1 WT & $-127.6 $ & $32.78$ &  $128.8$ &   $292.36$ \\\hline\hline
   & MM2 WT &  $ -280.98$ &$23.74$ &  $280.79$ &  $304.6$ \\\hline
     \\\hline\hline
   & TP06 & $-312.74$ & $39.82$ &  $373.65$ &   $302.12$ \\\hline\hline
$\bf{1000}$ & MM1 WT & $-144.46$ & $36.58$ &  $146.18$ &   $302.64$ \\\hline\hline
   & MM2 WT & $-300.43 $ & $ 25.11 $ & $300.49$ &  $314.84$ \\\hline
\end{tcolorbox}

\caption{\label{table1} \small \textbf{Characteristic properties of the action
potentials in TP06, MM1 WT, and MM2 WT models.} These data are for three
representative cases: low-frequency (PCL$=1000 ms$), intermediate-frequency
(PCL$=650 ms$), and high-frequency (PCL$=300 ms$) pacing.}

\label{Table2}
\end{table}

\end{document}